\documentclass{article}

\usepackage[preprint]{neurips_2025}
\usepackage{booktabs} % For professional looking tables
\usepackage{geometry} % To adjust page margins if needed
\usepackage{hyperref}
\hypersetup{
    unicode=false,          % non-Latin characters in Acrobat’s bookmarks
    colorlinks=true,        % false: boxed links; true: colored links
    linkcolor=blue,          % color of internal links (change box color with linkbordercolor)
    citecolor=blue,        % color of links to bibliography
    filecolor=magenta,      % color of file links
    urlcolor=cyan           % color of external links
}

\usepackage[utf8]{inputenc} % allow utf-8 input
\usepackage[T1]{fontenc}    % use 8-bit T1 fonts
\usepackage{url}            % simple URL typesetting
\usepackage{booktabs}       % professional-quality tables
\usepackage{amsfonts}       % blackboard math symbols
\usepackage{nicefrac}       % compact symbols for 1/2, etc.
\usepackage{microtype}      % microtypography
\usepackage{adjustbox}      % To center the table
\usepackage{lmodern} % Or another font package 
\usepackage{amsmath}
\usepackage{array}
\usepackage{tabularx}
\usepackage{tikz}
\usepackage{amsmath}
\usepackage{graphicx}
\usepackage{caption}
\usepackage[most]{tcolorbox} % For enhanced colored boxes
\usepackage{parskip} % Optional: for space between paragraphs instead of indent
\usepackage{multirow}
\usepackage{subfigure}

\usetikzlibrary{shapes, arrows, positioning, patterns, arrows.meta}

\tikzstyle{vec} = [rectangle, draw, minimum height=0.6cm, minimum width=0.2cm, fill=blue!20]
\tikzstyle{vec_pruned} = [rectangle, draw, minimum height=0.6cm, minimum width=0.2cm, fill=red!40]
\tikzstyle{vec_selected} = [rectangle, draw, minimum height=0.6cm, minimum width=0.2cm, fill=green!40]
\tikzstyle{cluster_node} = [circle, draw, fill=yellow!50, minimum size=0.5cm]
\tikzstyle{line} = [draw, -latex']

\definecolor{clustercolor0}{rgb}{0.121,0.466,0.705}
\definecolor{clustercolor1}{rgb}{1.000,0.498,0.055}
\definecolor{clustercolor2}{rgb}{0.172,0.627,0.172}
\definecolor{clustercolor3}{rgb}{0.839,0.152,0.156}

\newcommand{\crisp}{\textsc{CRISP}}
\title{\crisp: Clustering Multi-Vector Representations for Denoising and Pruning}

\author{%
  João Veneroso \\
  Google \\
  \texttt{jmfveneroso@google.com} \\
  \And
   Rajesh Jayaram \\
   Google Research \\
   \texttt{rkjayaram@google.com} \\
   \And
   Jinmeng Rao \\
   Google Deepmind \\
   \texttt{jinmengrao@google.com} \\
   \And
   Gustavo Hern{\'a}ndez {\'A}brego \\
   Google Deepmind \\
   \texttt{gustavoha@google.com} \\
   \And
   Majid Hadian \\
   Google Deepmind \\
  \texttt{majidh@google.com} \\
   \And
   Daniel Cer \\
   Google Deepmind \\
  \texttt{cer@google.com} \\
}

\begin{document}

\maketitle

\begin{abstract}
%In recent years, multi-vector retrieval has emerged as a increasingly important method of Neural Information Retrieval (IR).  Beginning with ColBERT, Multi-vector models have achieved state-of-the-art performance by representing documents by a collection of vectors, one per token, rather than a single vector. However, this performance comes with significant storage and computing costs due to the increased representation size, which is the primary bottleneck obstructing widespread adoption. To date, the only approach to pruning the number of vectors is to cluster them after generation. 
%In this work, we introduce XYZ, a novel multi-vector training method that learns pruned representations during end-to-end training. By incorporating the clustering step during training, XYZ learns to generate representations that are inherently clusterable, rather than imposing structure post-hoc, and allowing for for an arbitrary number of vectors to be output. We demonstrate that our method significantly outperforms clustering-post hoc at all scales of representation size, in addition to outperforming other baselines such as k-token-pooling. Importantly, our pruned representations \textit{outerperform} the original un-pruned representations for BEIR datasets with longer queries -- by learning good clusterings, our model filters irrelevant information and generates more accurate representations. 

Multi-vector models, such as ColBERT, are a significant advancement in neural information retrieval (IR), delivering state-of-the-art performance by representing queries and documents by multiple contextualized token-level embeddings. 
However, this increased representation size introduces considerable storage and computational overheads which have hindered widespread adoption in practice. A common approach to mitigate this overhead is to cluster the model's frozen vectors, but this strategy's effectiveness is fundamentally limited by the intrinsic clusterability of these embeddings.
 In this work, we introduce \crisp\ (Clustered Representations with Intrinsic Structure Pruning), a novel multi-vector training method which learns inherently clusterable representations directly within the end-to-end training process. By integrating clustering into the training phase rather than imposing it post-hoc, \crisp\ significantly outperforms post-hoc clustering at all representation sizes, as well as other token pruning methods.
 On the BEIR retrieval benchmarks, \crisp\ achieves a significant rate of \textbf{~3x} reduction in the number of vectors \textit{while outperforming} the original unpruned model. This indicates that learned clustering effectively denoises the model by filtering irrelevant information, thereby generating more robust multi-vector representations. With more aggressive clustering, \crisp\ achieves an \textbf{11x} reduction in the number of vectors with only a $3.6\%$ quality loss.
 %On the BEIR retrieval benchmarks, \crisp\ achieves an \textbf{11x} reduction in the number of vectors with only a $3.6\%$ quality loss. Surprisingly, when configured to compress at a still significant rate of \textbf{~3x}, \crisp's pruned representations actually \textit{outperform} the original unpruned model. This indicates that learned clustering effectively denoises the model by filtering irrelevant information, thereby generating more robust multi-vector representations.

%Importantly, \crisp's pruned representations outperform the original unpruned vectors on the BEIR evaluation tasks with longer queries, as the learned clustering filters out irrelevant information and generates more robust multi-vector representations.
 
 %\raj{Add concrete numbers to abstract.}
\end{abstract}

\section{Introduction}
\label{sec:introduction}

Neural embedding models are by now a foundational tool for representing data that underlie the SOTA methods for information retrieval (IR)~\cite{zhang2016neural}, clustering,  classification~\cite{muennighoff2022mteb}, among many other tasks. 
Recently, \emph{multi-vector} (MV) representations, introduced by the \textit{late-interaction} framework in ColBERT~\cite{khattab2020colbert}, have been shown to deliver significantly improved performance on popular IR benchmarks. ColBERT and its variants \cite{gao2021coil,hofstatter2022introducing,lee2024rethinking,lin2024fine,qian2023multivector,santhanam-etal-2022-colbertv2,wang2021pseudo,yao2021filip} produce \textit{multiple} embeddings per query or document by generating one embedding per token.  The similarity between a query and a document is calculated via so-called \emph{Chamfer Similarity}, also known as the MaxSim operation, between the corresponding sets of vectors. This enables a more fine-grained and expressive representation compared to using a single embedding per query or document, in addition to enabling improved
interpretability \cite{formal2021white,wang2023reproducibility} and generalization \cite{lupart2023ms,formal2022match,zhan2022evaluating,weller2023nevir}.

%\foootnote{ ColBERTv2 is now one of the most downloaded models on the HuggingFace model hub, with over 5million monthly downloads. See https://huggingface.co/colbert-ir/colbertv2.0}

 Despite these advantages, multi-vector representations are inherently more expensive than single-vector representations. Namely, producing embeddings for every input token increases representation size by multiple orders of magnitude. For instance, on the popular MS MARCO \cite{nguyen2016ms} dataset, the canonical ColbertV2 model \cite{santhanam-etal-2022-colbertv2} produces nearly 80 vectors per document and 32 per query. This increase in scale has many downstream effects. Firstly, it increases the storage requirements for indices which store multi-vector embeddings. Secondly, unlike inner product which is used for single-vector embeddings and has a linear cost in the embedding dimensionality, the runtime of the non-linear Chamfer Similarity scales \emph{quadratically} in the number of embeddings. As a result, the cost and quality of multi-vector retrieval algorithms depend heavily on the number of embeddings
  \cite{santhanam2022plaid, dhulipala2024muveramultivectorretrievalfixed}.

 The high cost for employing multi-vector embeddings has been a significant barrier to their widespread adoption. Thus, there has been considerable work in recent years to improve the efficiency of multi-vector models \cite{santhanam-etal-2022-colbertv2, engels2024dessert, hofstatter2022introducing, qian2023multivector, dhulipala2024muveramultivectorretrievalfixed,clavie2024reducingfootprintmultivectorretrieval,macavaney2025efficientconstantspacemultivectorretrieval}. For instance, the Colbertv2 \cite{santhanam-etal-2022-colbertv2, santhanam2022plaid} system and successors employ aggressive centroid-based quantization strategies to reduce index sizes, at the cost of significant complexity and tuning challenges \cite{macavaney2024reproducibility}. An alternative and enticing direction is to \textit{reduce} the number of vectors produced by the model entirely. Several such ``pruning'' methods have been studied, such as removing vectors which are not as significant to the overall query or passage; this can be done either be learned importance scores or gates \cite{hofstatter2022introducing} or by post-hoc pruning of individual vectors based on other mechanisms such as attention weights \cite{liu2024analysis}.
 
 While token-level pruning can be helpful, it has the downside of completely  dropping information which fails to exceed a given relevance threshold. Instead, one could hope to learn a holistic representation of the full data with fewer vectors directly. 
 A partial step towards this has been the technique of clustering the multi-vector representations post-hoc, after the model is trained \cite{clavie2024reducingfootprintmultivectorretrieval, dhulipala2024muveramultivectorretrievalfixed}. Each cluster is assigned a single pooled vector, which represents the cluster on the aggregate. Unfortunately, this post-hoc approach is limited by the actual clusterability of the frozen embeddings, which were not themselves trained to be clusterable.

 %These techniques have made notable progress, but still the best MV indices are nearly an order of magnitude larger than their single-vector counter-parts. Importantly, these techniques are only applied post-hoc, that is on top of a frozen multi-vector model; thus, they are limited to compressing the fixed output of such a model, rather than learning compressed multi-vector embeddings to begin with.  

%Perhaps change this following paragraph to be similar to the flow of the intro of https://arxiv.org/pdf/2409.14683
In this work, we look beyond post-hoc clustering of multi-vector representations, and study the \textit{learnability} of clustered multi-vector representations. In essence, we question the widespread assumption that token-level embeddings are necessary for the improved expressability of multi-vector models, and instead posit that in fact multi-vector models may have the potential to retain significant expressive capabilities representing data by clustered representations of their token-level embeddings. Specifically, we consider the question:

\begin{quote}
    \centering
    {\it Can multi-vector models be trained to produce inherently clusterable representations with negligible quality loss?
    }
\end{quote}

\subsection{Contributions}
%Mention Gemma somewhere?

We introduce \crisp\ (Clustered Representations with Intrinsic Structure Pruning), a novel training paradigm for multi-vector retrieval models. Unlike prior post-hoc clustering methods that operate on pre-trained, frozen embeddings, \crisp\ integrates clustering directly into the end-to-end training process. As a result, the model is trained to produce inherently clusterable representations, enabling a significant reduction in representation size with minimal impact on quality, and in some cases even leading to performance improvements. Our main contributions are summarized below. 

 \begin{itemize}
    \item \textbf{Significant Compression:} We demonstrate that \crisp's significantly improves representation size with minimal drops in quality. Evaluating on the BEIR retrieval benchmark \cite{thakur2021beir}, our \texttt{C8x32} \crisp\ model surpasses the unpruned multi-vector baseline by 0.4\%  while compressing document representations by \textbf{2.9x} and query representations by \textbf{3.9x}. Our more aggressive \texttt{C4x8} variant achieves a compression rate of \textbf{11x} for documents and \textbf{7.9x} for queries with only a 3.6\% drop in quality. 

    \item \textbf{Denoising Effect:} In addition to compressing representations, we show that \crisp\ acts as an effective denoising mechanism. By guiding the model to consolidate semantic information by training over clustered representations, \crisp\ learns to filter out less relevant token-level details, thereby generating more robust representations on datasets prone to noise. For instance, our \texttt{C4x8} model outperforms the unpruned model by 5.5\% on ArguAna, 6.8\% on Scidocs, and 2.7\% on NQ, all while reducing the number of document embeddings by 11x. In fact, averaged over all BEIR datasets, our \texttt{C8x32} model achieved the top score, \textit{outperforming} even the unpruned model (54.5 vs. 54.3 NDCG@10). 
    
    %This often results in pruned models that not only match but, as seen with \texttt{C8x32} (54.5 NDCG@10 vs. 54.3 for unpruned), can outperform the original unpruned models, particularly on datasets with longer queries (e.g., FIQA, NQ, Quora), highlighting a key quality contribution.
\item \textbf{Superiority over Post-Hoc Clustering:} \crisp\ demonstrates a clear advantage over traditional post-hoc clustering techniques. To achieve parity with unpruned models, post-hoc clustering methods achieve only a 2x compression rate limited to \textit{documents only}, whereas our \texttt{C8x32} model achieves parity with superior 2.9x document \emph{and} 3.9x query compression rates. At higher compression levels \crisp's benefits are even more pronounced. Our \texttt{C4x8} variant, despite its dramatic 11x document and 7.9x query compression rate, experiences only a 3.6\% drop in NDCG@10. This is a substantial improvement over post-hoc clustering, which reports a 9.3\% degradation limited to a 6x \textit{document-only} compression~\cite{clavie2024reducingfootprintmultivectorretrieval}. This underscores \crisp's more effective compression-quality trade-off, stemming from its end-to-end training.

    %\item \textbf{State of the Art:} Our comprehensive empirical evaluation on the BEIR benchmark, using a Gemma 2B backbone, shows that CRISP consistently and significantly outperforms post-hoc clustering techniques and other fixed pruning baselines. For instance, CRISP \texttt{C8x32} achieves a 0.4\% NDCG@10 improvement over an unpruned baseline with substantial document ($\sim$2.9x) and query ($\sim$3.9x) compression, surpassing post-hoc methods which might achieve a $\sim$0.6\% gain at only $\sim$2x \textit{document-only} compression [1]. Moreover, CRISP \texttt{C4x8} (with $\sim$11x document and $\sim$7.9x query compression) retains quality far better (only -3.6\% NDCG@10) than post-hoc approaches that see 3\% to 9.3\% degradation at 4x to 6x \textit{document-only} compression.
\end{itemize}

%\paragraph{Contributions.} In this work, we propose \crisp, a novel clustering-based training algorithm for multi-vector models, which significantly reduces representation size while maintaining high performance for retrieval tasks. CRISP achieves effective vector pruning by ultimately using clustered representations. But unlike post-hoc approaches, CRISP integrates clustering directly into its end-to-end training. This training process is designed to leverage and enhance the intrinsic structure, the inherent semantic groupings within the token embeddings, guiding the model to learn representations inherently optimized for clustering. The resulting reduction in vectors significantly lowers storage and computational overhead, often preserving or even improving retrieval accuracy through this structure-aware, learned denoising.

% TODO(majidh)

\section{Methodology}
\label{sec:method}

%\raj{Add Chamfer Similarity Equation}

This section details our proposed \crisp\ (Clustered Representations with Intrinsic Structure Pruning) methodology and the baseline pruning techniques against which it is compared. We begin by outlining the challenges with standard multi-vector representations that motivate this work, followed by a description of our base model architecture and the experimental approach to pruning.

\subsection{Background and Limitations of Multi-Vector Representations}
\label{subsec:limitations}

Multi-vector (MV) models learn more expressive representations than traditional single vector models by computing one embedding \textit{per-token} of the input text. This encodes queries and documents as \textit{sets} of vectors $Q,D \subset \mathbb{R}^d$ respectively. Multi-vector then scores the query-document similarity via the Chamfer Similarity \cite{dhulipala2024muveramultivectorretrievalfixed} (also known as MaxSim~\cite{khattab2020colbert}): 
\begin{equation}\label{eqn:chamfer}
     \textsc{Chamfer}(Q,D) = \sum_{q \in Q} \max_{x \in D} \langle q,x\rangle  
\end{equation}
where $\langle \cdot, \cdot \rangle$ is the standard inner product. Beginning with ColBERT~\cite{khattab2020colbert}, these multi-vector models have been shown to achieve significant performance improvements over single-vector models. However, due to their increased representation size, there are several key challenges associated with multi-vector models: 
\begin{itemize}
    \item \textbf{Computational Expense:} The increased number of vectors per item, combined with the Chamfer scoring that scales quadratically with the number of vectors ($O(M N d)$ for $M$ query and $N$ document vectors), makes MV models computationally expensive and significantly increases their memory footprint.
        \item \textbf{Semantic Redundancy and Skewing Similarity:} Repetitive tokens with similar contextual meaning in the queries can disproportionately affect the Chamfer similarity score, since embeddings of these tokens will appear multiple times in the similarity computation (\ref{eqn:chamfer}). This phenomenon is likely undesirable, since query-document similarity should ideally be based on distinct concepts in the text.
    \item \textbf{Information Noise:} Allocating uniform representational capacity (bits of information) to all tokens, irrespective of their semantic richness, can introduce noise and degrade performance compared to single-vector models in some cases, especially for datasets with long queries (e.g. ArguAna). By clustering of representations, one can hope to mitigate these issues by representing entire regions of the latent space by a single embedding, reducing the effect of outliers and sparsely related embeddings. 
\end{itemize}

Our work, \crisp, makes strides to alleviate all of the above issues by \textit{end-to-end} learning clustered multi-vector representations, thereby \textbf{(1)} significantly reducing the number of vectors, improving both computational costs and memory footprint, \textbf{(2)} diminishing the effect of redundant query tokens on the similarity, and \textbf{(3)} denoising representations by guiding the model towards representations without outliers or spread-clusters.

\subsection{Experimental Framework}
\label{subsec:framework}

\noindent\textbf{Base Model Architecture:}\\
The multi-vector models in this work utilize a dual encoder architecture with a \texttt{Gemma2B} backbone\cite{gemmateam2024gemmaopenmodelsbased}. We finetune \texttt{Gemma2B} with the Chamfer similarity loss~(\ref{eqn:chamfer}), using the standard methodology employed by ColBERT and models based on ColBERT~\cite{khattab2020colbert, santhanam-etal-2022-colbertv2}.
Building query and document encoders by leveraging pre-trained large language models (LLMs) as encoders has proven to be a strategy that produces high-performing embeddings for textual inputs \cite{Ni2022LargeDE, lee2024rethinking}.
As in \cite{khattab2020colbert} and \cite{lee2024rethinking}, no aggregation or pooling was utilized, and thus our model generates token-level representations.
This way, for each token in the textual input the model produces one dense vector representing it.
We chose not to project down our representations and kept each vector at the original dimension of 2048.
We use contrastive learning based on Chamfer similarity using the large collection of training datasets from \cite{li2025making} that includes publicly available data for retrieval, re-ranking, classification, clustering and sentence-to-sentence similarity (STS).
Our training setup includes in-batch random negatives, plus we used the hard negatives included in the retrieval training datasets.

%This setup generates multiple contextualized embeddings per token for both queries and documents, akin to the approach popularized by ColBERT. This allows for fine-grained interactions during scoring, moving beyond single-point comparisons of traditional dual encoders.

%\raj{Mention any other differences between our training process (Gemma) and ColBERT here! E.g. we don't cap query tokens at 32.}

%The multi-vector model utilized in this study employs a dual encoder architecture with a Gemma 2B backbone. While specific implementation details may vary, the core concept of generating multiple, contextualized embeddings per token for both query and document largely mirrors the approach popularized by models like ColBERT \cite{khattab2020colbert, santhanam-etal-2022-colbertv2}. This architecture facilitates fine-grained interactions between the query and document representations during the scoring phase, moving beyond the single-point comparison inherent in traditional dual encoders.

\noindent\textbf{Training over Pruned Representation}\\
We will compare \crisp, our clustering-based multi-vector training method, against several fixed-token pruning methods. For all pruned models we consider, the corresponding pruning strategy is applied \emph{during} training of the embeddings; i.e., if we prune to consider only the last $4$ tokens, then only these tokens are used in when computing the Chamfer loss during training. This ensures that the model training is aligned with its evaluation. Note that this would not be the case in, for instance, post-hoc clustering of an unpruned model.

%Applying pruning only at inference would create a train-test mismatch, hindering the model's ability to adapt its representations and potentially leading to suboptimal performance.

%focuses on integrating clustering directly into the model's training process to learn pruned, inherently structured representations.

%Pruning aims to address these challenges by reducing the number of vectors used during retrieval and scoring, potentially improving efficiency and acting as a denoising mechanism. In this paper, we introduce and evaluate \crisp, an approach focused on integrating clustering during training. We compare \crisp\ (implemented via K-means during training and inference) against fixed selection pruning methods applied to a Gemma 2B \cite{gemmateam2024gemmaopenmodelsbased} multi-vector model trained on the BGE dataset []. The methods fall into two main categories: fixed selection and clustering.

%Crucially, for the pruned models evaluated, the respective pruning strategies (Tail Pruning, K-Spacing, or Clustering) were applied consistently during both the training and inference stages. This ensures that the model learns to optimize its token embeddings and subsequent scoring based on the specific subset of vectors available post-pruning. Applying pruning only at inference would create a train-test mismatch, hindering the model's ability to adapt its representations and potentially leading to suboptimal performance.

The pruning strategies explored in this paper fall into two main types: fixed selection methods, which apply predefined heuristics, and clustering-based approaches, which group vectors by semantic similarity. We now detail the specific strategies within each category.

\subsection{Fixed-Token Pruning}
\label{subsec:fixed_token_pruning}
We first describe several approaches that select a fixed subset of token vectors using predefined, content-independent heuristics. In all cases, the model is trained with the loss function computed \textit{only} over these selected token vectors, tasking it with learning expressive representations under these constraints.

\subsubsection{Tail Pruning}
\label{subsubsec:tail_pruning}

%\raj{This seems to miss a major point: the model is trained on a loss which only includes the last 4 vectors. This is very different from just taking the last 4 or 8 vectors. The model therefore can in theory learn to cluster the representations and average them into those last 4 vectors, but this learning task is more difficult, and likely it will be harder for the model to learn how to correctly reorganize all the information into the last few vectors}

%   \raj{mention here that the model is trained over the last k tokens, i.e. pruning is performed during training, not post-hoc}

This method selects only the final $k_q$ token vectors from the query's sequence representation and $k_d$ from the document's, where $k_q$ is usually smaller than $k_d$. The main considerations for this method are twofold: first, for a pre-trained autoregressive language model, the later tokens might capture more summary information of the whole sequence; thus, keeping the later ones might better balance the performance-efficiency trade-off. Second, the query is usually much shorter than the document, and we can choose a smaller $k_q$ for the query but a larger $k_d$ for the document. Since the model is trained over the selected tokens, the goal is for the model to ``learn'' to move the most relevant information in the text into the embeddings for the last $k_q$ or $k_d$ tokens.
We tested two configurations:

\begin{itemize}
    \item \textbf{Tail Pruning (4x8)}: Selects the last 4 query vectors and last 8 document vectors.
    \item \textbf{Tail Pruning (8x32)}: Selects the last 8 query vectors and last 32 document vectors.
\end{itemize}
If there are fewer tokens in the input than the above fixed size, e.g. less than $k_q$ query tokens or less than $k_d$ document tokenss, then we simply select all tokens to train over.

%\raj{what about when there are less than 4 or 8 tokens? Do we take max(Seq length, 8) tokens)? Same for k-clustering below}

\begin{figure}[hbt!]
    \centering
    \begin{tikzpicture}[node distance=0.05cm]
        % --- Define Styles ---
        % Style for initial vectors (not retained in this pruning method): white fill, black border
        \tikzstyle{vec} = [rectangle, draw=black, minimum height=0.5cm, minimum width=0.25cm, fill=white, inner sep=1pt]
        % Style for tail vectors (retained): black fill, black border
        \tikzstyle{vec_tail} = [rectangle, draw=black, minimum height=0.5cm, minimum width=0.25cm, fill=black, inner sep=1pt]

        % --- Draw Vectors ---
        % Draw the vectors that are NOT part of the tail (first 6 in this example)
        \node (v1) [vec] {};
        \foreach \x in {2,...,6}{
            \node (v\x) [vec, right=of v\the\numexpr\x-1\relax] {};
        }
        % Draw the vectors that ARE part of the tail (last 4 in this example) using the updated style
        \node (v7) [vec_tail, right=of v6] {}; % Start the tail sequence
        \foreach \x in {8,...,10}{
            \node (v\x) [vec_tail, right=of v\the\numexpr\x-1\relax] {};
        }

        % --- Optional Labels ---
        \node [left=0.3cm of v1] {Embedding Sequence:};
        \node [below=0.3cm of v5.south] {(Example: Keeping last 4 vectors)};

    \end{tikzpicture}
    \caption{Illustration of Tail Pruning, where the last $k$ vectors (here $k=4$) are retained (shown in black), and others are shown in white.}
    \label{fig:tail_pruning_black_white}
\end{figure}

\subsubsection{K-Spacing}
This method uniformly subsamples the token vectors by selecting every $K$-th vector in the sequence. A key assumption for this method is that adjacent tokens might exhibit similar feature patterns. Consequently, we might perform pruning by sampling at a fixed interval, ensuring the remaining tokens are still evenly distributed throughout the sequence. We tested:
\begin{itemize}
    \item \textbf{K-Spacing (k=4)}: Selects every 4th vector (25\% density).
    \item \textbf{K-Spacing (k=2)}: Selects every 2nd vector (50\% density).
\end{itemize}

\begin{figure}[hbt!]
    \centering
    \begin{tikzpicture}[node distance=0.05cm]
        % --- Define Styles ---
        % Style for unselected vectors: white fill, black border
        \tikzstyle{vec} = [rectangle, draw=black, minimum height=0.5cm, minimum width=0.25cm, fill=white, inner sep=1pt]
        % Style for selected spaced vectors: black fill, black border
        \tikzstyle{vec_spaced} = [rectangle, draw=black, minimum height=0.5cm, minimum width=0.25cm, fill=black, inner sep=1pt]

        % --- Draw Vectors ---
        % Draw vectors, alternating style for k=2
         \node (v1) [vec_spaced] {};
         \node (v2) [vec, right=of v1] {};
         \node (v3) [vec_spaced, right=of v2] {};
         \node (v4) [vec, right=of v3] {};
         \node (v5) [vec_spaced, right=of v4] {};
         \node (v6) [vec, right=of v5] {};
         \node (v7) [vec_spaced, right=of v6] {};
         \node (v8) [vec, right=of v7] {};
         \node (v9) [vec_spaced, right=of v8] {};
         \node (v10) [vec, right=of v9] {};

        % --- Optional Labels ---
        \node [left=0.3cm of v1] {Embedding Sequence:};
        \node [below=0.3cm of v5.south] {(Example: K-Spacing k=2)}; % Label for context
    \end{tikzpicture}
    \caption{Illustration of K-Spacing, where every $k$-th vector (here $k=2$) is retained (shown in black), and others are shown in white.}
    \label{fig:k_spacing_black_white}
\end{figure}

\subsection{Clustering-Based Pruning (The \crisp\ Approach)}
We now describe the clustering-based methods which constitute the \crisp\ concept. These methods use K-means clustering to group semantically similar token vectors, and then aggregate the clusters into a single vector by using the centroids (i.e., mean-pooling each cluster). By aggregating tokens with similar contextual meanings into a single cluster and representing them with their centroid, \crisp\ ensures that the contribution comes from distinct semantic units rather than multiple, near-identical token representations. This dynamic grouping also allows for a more flexible allocation of representational capacity, where densely packed semantic regions of the text are captured by individual cluster centroids, while less informative regions might contribute to larger, more general clusters, effectively focusing the model's attention on the most salient aspects of the text.

The key hyperparameter is the choice of $k$, which is the number of clusters and thus the number of vectors used to represent each query or document. We consider two methods of for selecting the hyperparameter $k$.

\subsubsection*{Fixed-Size Clustering}
K-means is applied to obtain a pre-defined number of clusters ($k_q$ for query, $k_d$ for document). We tested:
\begin{itemize}
    \item \textbf{Clustering (4x8)}: Uses $k_q=4$ query centroids and $k_d=8$ document centroids.
    \item \textbf{Clustering (8x32)}: Uses $k_q=8$ query centroids and $k_d=32$ document centroids.
\end{itemize}

\subsubsection*{Relative-Size Clustering}
K-means is applied, but the number of clusters $k$ is set relative to the original sequence length $L$. All resulting centroids are used in scoring. We tested:
\begin{itemize}
    \item \textbf{Clustering (25\%)}: $k = \left\lfloor 0.25 \times L \right\rfloor$.
    \item \textbf{Clustering (50\%)}: $k = \left\lfloor 0.50 \times L \right\rfloor$.
\end{itemize}

\begin{figure}[hbt!]
    \centering
    \begin{tikzpicture}[node distance=0.05cm and 0.5cm] % Adjust vertical and horizontal spacing
        % --- Define Styles ---
        % Style for Cluster 1 vectors: white fill, black border
        \tikzstyle{vec_cluster1} = [rectangle, draw=black, minimum height=0.5cm, minimum width=0.25cm, fill=white, inner sep=1pt]
        % Style for Cluster 2 vectors: black fill, black border
        \tikzstyle{vec_cluster2} = [rectangle, draw=black, minimum height=0.5cm, minimum width=0.25cm, fill=black, inner sep=1pt]
        % Style for Cluster 3 vectors: gray fill, black border
        \tikzstyle{vec_cluster3} = [rectangle, draw=black, minimum height=0.5cm, minimum width=0.25cm, fill=gray!60, inner sep=1pt]
        
        % Default vec style (not used in current example but defined for completeness)
        \tikzstyle{vec} = [rectangle, draw=black, minimum height=0.5cm, minimum width=0.25cm, fill=gray!30, inner sep=1pt]

        \tikzstyle{centroid} = [circle, draw, thick, minimum size=0.6cm, inner sep=1pt]
        \tikzstyle{arrow} = [draw, -{Stealth[length=2mm]}]

        % --- Draw Vectors ---
        \node (v1) [vec_cluster1] {};
        \node (v2) [vec_cluster1, right=of v1] {};
        \node (v3) [vec_cluster1, right=of v2] {};
        \node (v4) [vec_cluster1, right=of v3] {};
        
        \node (v5) [vec_cluster2, right=of v4] {};
        \node (v6) [vec_cluster2, right=of v5] {};
        \node (v7) [vec_cluster2, right=of v6] {};
        \node (v8) [vec_cluster2, right=of v7] {};
        
        \node (v9) [vec_cluster3, right=of v8] {};
        \node (v10) [vec_cluster3, right=of v9] {};
        \node (v11) [vec_cluster3, right=of v10] {};
        \node (v12) [vec_cluster3, right=of v11] {};
        \node[left=0.3cm of v1] (orig_label) {Original Vectors:}; % Label for the sequence

        % --- Draw Centroids ---
        % Centroid 1: White fill, black border
        \node (c1) [centroid, fill=white, below right=1.2cm and 0.15cm of v2.south] {C1};
        % Centroid 2: Black fill, black border, white text
        \node (c2) [centroid, fill=black, text=white, right=2.46cm of c1] {C2};
        % Centroid 3: Gray fill, black border
        \node (c3) [centroid, fill=gray!60, right=2.46cm of c2] {C3};
        \node[left=0.3cm of c1] {Centroids:}; % Label for centroids

        % --- Draw Mapping Arrows ---
        % Draw curved arrows from vectors to their assigned centroid
        \foreach \i in {1,...,4}
             \draw [arrow] (v\i.south) .. controls +(south:0.6) and +(north:0.6) .. (c1.north);
        \foreach \i in {5,...,8}
             \draw [arrow] (v\i.south) .. controls +(south:0.6) and +(north:0.6) .. (c2.north);
        \foreach \i in {9,...,12}
             \draw [arrow] (v\i.south) .. controls +(south:0.6) and +(north:0.6) .. (c3.north);

    \end{tikzpicture}
    \caption{Illustration of Clustering Pruning: Original vectors (top row, styled by assigned cluster) are mapped to their respective cluster centroids (bottom row). Here, 3 clusters (C1-C3) are used, distinguished by white, black, and gray fills. The selected embeddings for each cluster are calculated dynamically so they need not be adjacent.}
    \label{fig:clustering_detailed_accessible_gray}
\end{figure}
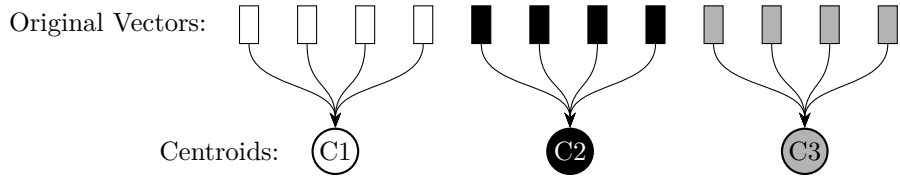

These methods were evaluated against the baseline single-vector and full multi-vector models, as detailed in the Experiments section.

\section{Experiments}
\label{sec:experiments}

\begin{table}[!htbp]
  \centering  
  %\caption{Dataset Statistics for Candidate and Query Tokens with Reduction Metrics and Thresholds}  
  \label{tab:all_token_stats_final}

  \begin{subtable}
    \centering
    \small  
    \caption{Candidate Representation Size (relative to full MV)}
    \label{tab:candidate_data_stacked_final}
    \begin{tabular}{@{}l|r|rr cc@{}} 
    \toprule
    BEIR Task &  \begin{tabular}{@{}c@{}}Avg Candidate \\ Tokens \end{tabular} & \begin{tabular}{@{}c@{}} Relative Size \\  \texttt{4x8}, \texttt{C4x8}\end{tabular} & \begin{tabular}{@{}c@{}} Relative Size  \\  \texttt{8x32}, \texttt{C8x32} \end{tabular} & \begin{tabular}{@{}c@{}} Relative Size  \\ \texttt{K2}, C50 \end{tabular}  & \begin{tabular}{@{}c@{}} Relative Size  \\ K4, C25 \end{tabular}\\
    \midrule
    arguana & 206.68 & 0.039 & 0.155 & $<.50$ & $<.25$ \\
    climate\_fever & 142.59 & 0.056 & 0.224 & $<.50$ & $<.25$ \\
    dbpedia\_entity & 96.01 & 0.083 & 0.333 & $<.50$ & $<.25$ \\
    fever & 142.59 & 0.056 & 0.224 & $<.50$ & $<.25$ \\
    fiqa & 165.34 & 0.048 & 0.194 & $<.50$ & $<.25$ \\
    hotpotqa & 77.32 & 0.103 & 0.414 & $<.50$ & $<.25$ \\
    msmarco & 80.74 & 0.099 & 0.396 & $<.50$ & $<.25$ \\
    nfcorpus & 331.29 & 0.024 & 0.097 & $<.50$ & $<.25$ \\
    nq & 97.28 & 0.082 & 0.329 & $<.50$ & $<.25$ \\
    quora & 14.75 & 0.542 & 2.169 & $<.50$ & $<.25$ \\
    scidocs & 211.06 & 0.038 & 0.152 & $<.50$ & $<.25$ \\
    scifact & 302.06 & 0.026 & 0.106 & $<.50$ & $<.25$ \\
    trec\_covid & 226.36 & 0.035 & 0.141 & $<.50$ & $<.25$ \\
    webis\_touche2020 & 221.07 & 0.036 & 0.145 & $<.50$ & $<.25$ \\
    \midrule 
    \textbf{Average} & \textbf{165.37} & \textbf{0.091} & \textbf{0.349} &\textbf{0.50} &  \textbf{0.25}\\
    \bottomrule
    \end{tabular}
  \end{subtable}

  \vspace{2em}  

  \begin{subtable}
    \centering
    \small  
    \caption{Query Representation Size (relative to full MV)}
    \label{tab:query_data_stacked_final_mod} 
    \begin{tabular}{@{}l|r|rr cc@{}} 
    \toprule
    BEIR Task &  \begin{tabular}{@{}c@{}}Avg Query \\ Tokens \end{tabular} & \begin{tabular}{@{}c@{}} Relative Size \\ for \texttt{4x8}, \texttt{C4x8}\end{tabular} & \begin{tabular}{@{}c@{}} Relative Size  \\  \texttt{8x32}, \texttt{C8x32}\end{tabular} & \begin{tabular}{@{}c@{}} Relative Size  \\ \texttt{K2}, C50 \end{tabular}  & \begin{tabular}{@{}c@{}} Relative Size  \\ K4, C25 \end{tabular} \\
    \midrule
    arguana & 255.48 & 0.016 & 0.031 & $<.50$ & $<.25$ \\
    climate\_fever & 44.31 & 0.090 & 0.181 & $<.50$ & $<.25$ \\
    dbpedia\_entity & 23.75 & 0.168 & 0.337 & $<.50$ & $<.25$ \\
    fever & 28.44 & 0.141 & 0.281 & $<.50$ & $<.25$ \\
    fiqa & 31.40 & 0.127 & 0.255 & $<.50$ & $<.25$ \\
    hotpotqa & 39.88 & 0.100 & 0.201 & $<.50$ & $<.25$ \\
    msmarco & 25.21 & 0.159 & 0.317 & $<.50$ & $<.25$ \\
    nfcorpus & 21.41 & 0.187 & 0.374 & $<.50$ & $<.25$ \\
    nq & 25.97 & 0.154 & 0.308 & $<.50$ & $<.25$ \\
    quora & 31.32 & 0.128 & 0.255 & $<.50$ & $<.25$ \\
    scidocs & 33.31 & 0.120 & 0.240 & $<.50$ & $<.25$ \\
    scifact & 35.98 & 0.111 & 0.222 & $<.50$ & $<.25$ \\
    trec\_covid & 29.82 & 0.134 & 0.268 & $<.50$ & $<.25$ \\
    webis\_touche2020 & 25.86 & 0.155 & 0.309 & $<.50$ & $<.25$ \\
  \midrule 
    \textbf{Average} & \textbf{46.58} & \textbf{0.128} & \textbf{0.256} &  \textbf{0.50} &  \textbf{0.25}\\
    \bottomrule
    \end{tabular}
  \end{subtable}
  \vspace{1 em}
  \caption*{Tables \ref{tab:candidate_data_stacked_final} and \ref{tab:query_data_stacked_final_mod} show the average relative representation size, in terms of number of output embeddings, for each pruned multi-vector model compared to the unpruned model. Thus, the unpruned model has a relative size of 1, whereas $\texttt{C50}$, which outputs half as many of the embeddings, will have a relative size of at most $50\%$. An average relative size of $.091$ thus corresponds to a compression rate of $1/.091 \approx 11$. }
\end{table}

\subsection{Experimental Setup}

%\raj{what is the embedding dimension}
%\gustavo{added a note about this in Base Model Architecture: PTAL}

%\raj{Change 4x8, C8x32 to \texttt{4x8}. \texttt{C8x32} for emphasis}

%\raj{TODO: (Raj will do) add explanation of the new tables}

%\raj{mention that XTR uses a different model.}

We evaluate \crisp\, and compare its cluster-based training mechanism with the pruning methods described in Section \ref{subsec:fixed_token_pruning}.  All multi-vector and single-vector models, including the pruned variants, were built upon the \texttt{Gemma2B} pre-trained LLM. The models were fine-tuned for 20,000 steps using a batch of size 128 on Cloud TPU v3 \footnote{https://cloud.google.com/tpu/docs/v3} using the BGE\footnote{https://huggingface.co/datasets/cfli/bge-full-data/tree/main/data} dataset~\cite{li2025making}. Crucially, the pruned multi-vector variants (\texttt{4x8}, \texttt{8x32}, \texttt{K4}, \texttt{K2}, \texttt{C4x8}, \texttt{C8x32}, \texttt{C25}, \texttt{C50}) were fine-tuned starting from the multi-vector (`MV') baseline model. %All fine-tuning processes were conducted with a batch size of 128.

%This was repeated earlier
%Crucially, for the models employing pruning or clustering (representing the evaluated \crisp\ configurations), the corresponding strategy was consistently applied during both the training phase and subsequent inference to ensure the model learned representations optimized for the pruned vector space.

 We evaluate the performance of our models over all the BEIR~\cite{thakur2021beir} benchmark tasks, a standard suite for evaluating zero-shot retrieval performance. As it is standard for BEIR evaluations, we use NDCG@10 as our metric. We compare our pruned multi-vector models against the standard single-vector (`SV') and multi-vector (`MV') \texttt{Gemma2B} models. 
 In addition, when available, we report scores for two other models: GTR$_\text{xxl}$~\cite{Ni2022LargeDE}, which is a high-quality single-vector model, and XTR$_\text{xxl}$~\cite{lee2024rethinking}, which is a SOTA multi-vector model. Both models are trained from the same T5 backbone~\cite{raffel2020t5}, which differs from our \texttt{Gemma2B} backbone, and thus the performance numbers are not directly comparable. However, their inclusion serves to establish a reasonable performance baseline on BEIR.

 In all our multi-vector models, including the `MV' baseline and all pruned variants, inference was conducted using brute-force Chamfer similarity scoring comparing each query against every document in the corresponding dataset corpus. This ensures precise evaluation without confounding factors from approximate retrieval methods.

Given that the base \texttt{Gemma2B} model is instruction-tuned, adapting it for the varied information retrieval tasks within the BEIR benchmark required the consistent prepending of task-specific instruction prefixes to the input queries during both the training and inference stages. Initial experiments revealed a substantial decline in retrieval performance when these instructions were absent, suggesting that the model relies on this conditioning to align with the specific retrieval objectives of each task. The precise instruction prefixes utilized for each BEIR task are detailed in Appendix B.

\subsection{Results and Analysis}

The performance results for all models evaluated on the BEIR tasks are presented in Table~\ref{tab:model-performance-bold}. The table includes scores for our baseline single-vector (`SV') and multi-vector (`MV') models, the fixed selection pruning methods (Tail Pruning `\texttt{4x8}', `\texttt{8x32}'; K-Spacing `\texttt{K4}', `\texttt{K2}'), the clustering-based methods implementing the \crisp\ approach (`\texttt{C4x8}', `\texttt{C8x32}', `\texttt{C25}', `\texttt{C50}'), and the non-Gemma based models: XTR \cite{lee2024rethinking} and GTR~\cite{Ni2022LargeDE}. Tables \ref{tab:candidate_data_stacked_final} and \ref{tab:query_data_stacked_final_mod} show the size of each pruned multi-vector representation relative to the unpruned model for both query and document representations.

%\raj{sentence about XTR / GTR being somewhat incomparable here?}

\begin{table}[htbp]
  \caption{Pruning in BEIR (NDCG@10)} % Updated caption
  \label{tab:model-performance-bold} % Updated label
  \begin{adjustbox}{width=\textwidth, center}
  \begin{tabular}{l|rr|rr|rrrr|rrrr}
    \toprule
    & \multicolumn{2}{c|}{\multirow{3}{*}[0.5ex]{\textbf{External Baselines}}} & \multicolumn{10}{c}{\textbf{Gemma 2B-based Models}} \\
   \cmidrule(lr){4-13}
    \multirow{2}{*}[1.0ex]{\textbf{BEIR Task}} & & & \multicolumn{2}{|c}{\textbf{Unpruned}} & \multicolumn{4}{|c}{\textbf{Fixed Selection}} & \multicolumn{4}{|c}{\textbf{CRISP}}  \\
    \cmidrule(lr){2-3} \cmidrule(lr){4-13}
                      & XTR           & GTR           & MV            & SV            & \texttt{4x8}           & \texttt{8x32}          & K4            & \texttt{K2}            & \texttt{C4x8}          & \texttt{C8x32}         & C25           & C50           \\
    \midrule
    arguana                  & 44.2          & 54.0          & 70.1          & 73.8          & 53.6          & 56.4          & 67.4          & 69.3          & \textbf{74.0} & 71.6          & 66.4          & 69.2          \\
    climate\_fever           & 24.5          & 26.7          & 39.6          & 35.7          & 8.1           & 10.3          & 35.7          & \textbf{40.0} & 36.0          & 36.0          & 28.4          & 31.3          \\
    dbpedia\_entity          & \textbf{44.3} & 40.8          & 42.1          & 35.4          & 17.1          & 29.3          & 32.4          & 40.4          & 37.5          & 38.5          & 39.2          & 37.8          \\
    fever                    & 77.0          & 74.0          & \textbf{90.3} & 82.8          & 28.5          & 34.3          & 87.4          & 89.6          & 87.3          & 83.5          & 83.5          & 88.4          \\
    fiqa                     & 43.8          & 46.7          & 49.4          & 45.6          & 16.1          & 25.2          & 42.0          & 47.2          & 47.4          & \textbf{50.1} & 46.6          & 49.2          \\
    hotpotqa                 & 66.2          & 59.9          & \textbf{72.6} & 59.2          & 16.1          & 25.9          & 62.7          & 70.6          & 68.5          & 70.1          & 69.0          & 70.6          \\
    msmarco                  & \textbf{46.6} & 44.2          & 25.9          & 39.9          & 13.7          & 17.3          & 23.7          & 29.4          & 41.7          & 42.9          & 42.1          & 43.3          \\
    nfcorpus                 & 35.3          & 34.2          & \textbf{36.4} & 31.8          & 16.4          & 27.8          & 33.5          & 36.0          & 32.3          & 36.0          & 34.3          & 36.2          \\
    nq                       & 60.9          & 56.8          & 62.0          & 56.8          & 27.5          & 40.0          & 52.7          & 62.5          & 63.7          & \textbf{65.2} & 62.6          & \textbf{65.2} \\
    quora                    & 88.1          & 89.2          & 87.7          & 87.4          & 63.7          & 81.4          & 80.4          & 86.1          & 89.1          & \textbf{89.3} & 88.6          & 89.2          \\
    scidocs                  & 17.1          & 16.1          & 21.9          & 18.8          & 8.8           & 13.7          & 18.1          & 21.7          & 23.4          & 23.2          & 20.9          & \textbf{23.8} \\
    scifact                  & \textbf{74.3} & 66.2          & 73.8          & 49.5          & 21.6          & 36.3          & 65.6          & 72.3          & 55.5          & 65.8          & 45.7          & 58.9          \\
    trec\_covid              & \textbf{78.9} & 50.1          & 63.0          & 53.5          & 22.8          & 29.0          & 54.4          & 60.5          & 52.0          & 63.2          & 50.6          & 48.0          \\
    webis\_touche2020        & \textbf{30.9} & 23.3          & 25.0          & 28.8          & 8.5           & 18.6          & 22.0          & 24.6          & 23.7          & 27.3          & 25.6          & 24.4          \\
    \midrule
    Total                    & 52.7          & 49.1          & 54.3          & 49.9          & 23.0          & 31.8          & 48.4          & 53.6          & 52.3          & \textbf{54.5} & 50.3          & 52.5          \\
        Avg Rel Doc Size                    & ---          & ---          & 1          & ---        & .091          & .349          & .25       & .50          & .091          & .349 & .25        & .50          \\
          Avg Rel Query Size                    & ---          & ---          & 1          & ---        & .128          & .256          & .25       & .50          &  .128          & .256 & .25        & .50          \\
    \midrule
    cqadupstack              & ---           & ---           & 38.7          & 34.7          & 10.8          & 8.9           & 31.2          & 37.1          & 38.6          & \textbf{42.1} & 36.3          & 41.7          \\
    cq\_android     & ---           & ---           & 45.6          & 43.4          & 20.6          & 16.7          & 39.3          & 44.7          & 46.5          & \textbf{50.6} & 47.6          & 50.4          \\
    cq\_english     & ---           & ---           & 47.2          & 42.3          & 12.0          & 10.1          & 40.3          & 45.3          & 46.1          & \textbf{49.5} & 44.0          & 48.7          \\
    cq\_gaming      & ---           & ---           & 50.2          & 49.9          & 10.3          & 3.9           & 40.3          & 48.7          & 54.6          & 57.9          & 51.2          & \textbf{59.3} \\
    cq\_gis         & ---           & ---           & 34.4          & 28.2          & 6.4           & 4.5           & 25.8          & 32.6          & 34.7          & \textbf{39.8} & 31.9          & 37.8          \\
    cq\_mathematica & ---           & ---           & 28.7          & 24.4          & 9.9           & 8.5           & 22.7          & 28.0          & 26.2          & \textbf{31.8} & 23.5          & 29.3          \\
    cq\_physics     & ---           & ---           & 44.1          & 42.4          & 15.5          & 16.0          & 36.9          & 43.1          & 44.2          & 46.4          & 42.6          & \textbf{47.7} \\
    cq\_programmers & ---           & ---           & 41.0          & 39.6          & 9.8           & 7.4           & 35.8          & 40.1          & 41.4          & 42.9          & 38.0          & \textbf{43.4} \\
    cq\_stats       & ---           & ---           & 35.8          & 30.1          & 7.6           & 5.2           & 26.5          & 33.4          & 32.9          & 36.3          & 30.9          & \textbf{36.6} \\
    cq\_tex         & ---           & ---           & 27.6          & 22.0          & 6.2           & 5.2           & 20.0          & 25.7          & 26.0          & \textbf{30.3} & 22.0          & 27.9          \\
    cq\_unix        & ---           & ---           & 42.8          & 32.5          & 12.2          & 11.8          & 35.2          & 41.3          & 39.0          & \textbf{43.5} & 37.5          & 42.2          \\
    cq\_webmasters  & ---           & ---           & 39.7          & 36.4          & 13.8          & 13.3          & 32.9          & 37.8          & 38.9          & 41.6          & 38.1          & \textbf{43.2} \\
    cq\_wordpress   & ---           & ---           & 26.8          & 24.7          & 4.7           & 4.0           & 18.5          & 24.8          & 33.1          & \textbf{35.0} & 28.7          & 34.0          \\
    \bottomrule
  \end{tabular}
  \end{adjustbox}
\end{table}

Consistent with prior findings, the baseline `MV' model (Avg 54.3) significantly outperforms the baseline `SV' model (Avg 49.9), demonstrating the general advantage of multi-vector representations. Fixed selection pruning methods, however, lead to a substantial degradation in performance. Both Tail Pruning variants (`\texttt{4x8}': 23.0, `\texttt{8x32}': 31.8) and K-Spacing (`\texttt{K4}': 48.4, `\texttt{K2}': 53.6) score considerably lower than the `MV' baseline on average. Even `\texttt{K2}', which retains 50\% of the vectors, does not fully recover the baseline performance.

In contrast, the \crisp\ models prove much more effective. \crisp\ is significantly better than fixed selection in almost all BEIR tasks, particularly when compared to methods like `\texttt{4x8}' or `\texttt{8x32}' which consider only a small, fixed positional subset of vectors. The best performing model overall is the \crisp\  `\texttt{C8x32}' model (Avg 54.5), which slightly surpasses the performance of the full `MV' baseline. Other \crisp\ variants like `\texttt{C4x8}' (52.3) `C25' (50.2) and `C50' (52.5) also remain competitive and substantially outperform the fixed selection models. This suggests that K-means clustering effectively identifies and retains the most salient semantic information within the multiple token embeddings. Further, we observed that \crisp\ can indeed improve upon the vanilla multi-vector baseline in several cases (e.g., ArguAna `\texttt{C4x8}' vs `MV'; FIQA `\texttt{C8x32}' vs `MV'; NQ `\texttt{C8x32}'/`C50' vs `MV'; Quora `\texttt{C8x32}' vs `MV'), potentially due to a denoising effect where less informative token variations within clusters are abstracted away by the centroid representation (discussed in Appendix A). The overall strong performance of `\texttt{C8x32}' indicates that the \crisp\ approach provides a viable path towards reducing the complexity of multi-vector models while maintaining, and sometimes enhancing, retrieval effectiveness.

It is important to note that the effectiveness of all the models presented were found to be highly sensitive to hyperparameter tuning. Specifically, the learning rate and L2 weight normalization played critical roles in achieving the reported results. Furthermore, as detailed in Appendix \ref{sec:appendix_prefixes}, the practice of prepending task-specific instruction prefixes to the input queries during inference proved indispensable. Omitting or poorly configuring these elements significantly degraded retrieval performance, underscoring that the presented qualitative examples and quantitative successes are contingent upon careful optimization of these crucial training and inference parameters.

%Tables \ref{tab:candidate_data_stacked_final} and \ref{tab:query_data_stacked_final_mod} show the size of each pruned multi-vector representation relative to the unpruned model for both the query and document representations. The high performing \texttt{C4x8} model has a compression rate of $1/.091 \approx 10.98$, for queries and $1/.128 \approx 7.81$ for documents, and the top-performing \texttt{C4x8} achieves compression rates of $1/.349\approx2.87$ and $1/.256\approx 3.91$ for queries and documents respectively. Thus, the high performance of these models is complimented by a significant rate of compression.
 
\paragraph{Comparison to Post-Hoc Clustering.}
A primary alternative for comparison is \emph{post-hoc} clustering, where clustering is applied to a multi-vector model after its embeddings are frozen.  This approach offers a simpler training pipeline as it avoids clustering during the training phase, making it a natural baseline.  For example, post-hoc clustering was employed to enhance retrieval latency in \cite{dhulipala2024muveramultivectorretrievalfixed}, and is the main focus of \cite{clavie2024reducingfootprintmultivectorretrieval}, which explores clustering methods such as k-means and Hierarchical Clustering for pruning multi-vector representations. 
They found Hierarchical clustering to be the best performing method when evaluated over a subset of the BEIR retrieval tasks. Specifically, they showed that, compared to the unpruned  multi-vector model, this method yields: a 2x compression with a 0.6$\%$ NDCG@10 improvement, a 4x compression with a 3$\%$ NDCG@10 decrease, and a 6x compression with a 9.3$\%$ NDCG@10 decrease. Notably, in \cite{clavie2024reducingfootprintmultivectorretrieval} the authors applying pruning to the document representations only, leaving the query embeddings unchanged.

\crisp's cluster-based training, however, presents considerably more favorable compression-quality trade-offs.  The \texttt{C8x32} \crisp\ model, for instance, achieves \textbf{2.9x} document token compression, in addition to a \textbf{3.9x} query token compression, while improving NDCG@10 by 0.4$\%$.  Furthermore, the more aggressive \texttt{C4x8} \crisp\ model delivers \textbf{11x} document token and \textbf{7.9x} query token compression with only a 3.6$\%$ decrease in NDCG@10.  Therefore, \crisp\ not only yields better document compression rates with comparable or superior quality retention than post-hoc clustering, but crucially also enables substantial query-token compression simultaneously --- a benefit not achieved in \cite{clavie2024reducingfootprintmultivectorretrieval}. We emphasize that compressing query tokens has significant impact on downstream task, such as reducing retrieval latency, as demonstrated in \cite{dhulipala2024muveramultivectorretrievalfixed} where pre-clustering query tokens markedly sped up retrieval, and in \cite{santhanam2022plaid} where each query vector necessitates a separate vector-index query.

\section{Other Related Work}
\label{sec:related}

The seminal ColBERT model \cite{khattab2020colbert} introduced multi-vector models as a way to improve passage and document search via token-level interactions and representations of textual inputs.  
%ColBERT innovated in two relevant ways: demonstrated that the use of multi-vector representations was able to beat other alternatives for retrieval and re-ranking, and that leveraging pre-trained language models offered a clear advantage over models trained directly for retrieval.
%In that vein, the GTR model \cite{Ni2022LargeDE} leveraged the pre-trained encoder of the T5 model \cite{xue2021mt5} to build generalizable retrieval models and, despite using a single-vector representation, beat ColBERT in the BEIR \cite{thakur2021beir} retrieval evaluation.
%GTR demonstrated the importance of both the quality of the pre-trained model and its scale.Since then, several retrieval models have been built on top of more advanced pre-trained language models such as LLaMa \cite{ma2023finetuningllamamultistagetext}, Mistral \cite{lee2025nvembedimprovedtechniquestraining} and Gemini \cite{lee2025geminiembeddinggeneralizableembeddings}.
Since then, significant effort has been devoted to developing improved and optimized multi-vector models and retrieval methods \cite{santhanam-etal-2022-colbertv2,santhanam2022plaid,gao2021coil,hofstatter2022introducing,lee2024rethinking,lin2024fine,qian2023multivector,santhanam-etal-2022-colbertv2,wang2021pseudo,yao2021filip,  dhulipala2024muveramultivectorretrievalfixed}. Nevertheless, multi-vector models still require non-trivial computational overheads when compared to single-vector models. 
To address these efficiency issues, \cite{macavaney2025efficientconstantspacemultivectorretrieval} use a fixed number of vectors irrespective of the length of the queries or the documents and demonstrated that they are able to retain the performance of ColBERT-v2 \cite{santhanam-etal-2022-colbertv2} after reduction.
In contrast, \cite{lee2024rethinking} trained the multi-vector model to purposely use during retrieval the vectors that represent the most salient parts of the queries and the documents.
ALIGNER \cite{qian2023multivector} introduced linear programming for sparse query-document alignments and per token-salience.   
%In a different approach, \cite{clavie2024reducingfootprintmultivectorretrieval} addressed the trade-off by post-hoc clustering of the frozen embeddings.
%The vectors within the cluster are then averaged together using different pooling variants.
%Quantization is then applied to further reduce the size of the pooled vectors.
%Importantly, these two approaches do not adjust the model directly and the vector reduction operations are done post-hoc.
ColBERTer \cite{hofstatter2022introducing} followed up with a vector reduction approach based on pooling embeddings of tokens within the same word, as well as pruning by removing stop words. However, this pooling is fixed in advance and not learned based on the neural representations of the input tokens. In contrast to these prior approaches, \crisp\ is the first method to learn a pooling (i.e.  clustering) of the token representations during training time, allowing the model to couple the learning of the representations to the learning of the pooling.

%nudge the multi-vector representation towards a clust

%, showing that storage reduction is not at odds with multi-vector representation power.

%On the other hand, to address the added complexity of using multi-vector representations for retrieval \cite{dhulipala2024muveramultivectorretrievalfixed} introduce Fixed Dimensional Encodings (FDEs) of queries and documents whose inner product%approximates multi-vector similarity.

\section{Conclusion}

%\raj{Reminder to make pass on this section, add limitations}

In this paper, we introduced \crisp\ (Clustered Representations with Intrinsic Structure Pruning), a novel multi-vector model method that learns inherently clusterable representations during end-to-end training, thereby significantly reducing the representation size of multi-vector models. \crisp\ produces pruned models that actually \emph{outperform} the original unpruned models, while compressing document representations by \textbf{3x}, as well as offering an improved \textbf{11x} compression at the cost of a small drop in performance. By learning clusters during training, \crisp\ also significantly outperforms post-hoc clustering methods that operate on frozen embeddings. Thus, \crisp\ offers a significant step towards bridging the efficiency gap between multi-vector and single-vector models.

\textbf{Broader Impacts and Limitations:}

Our work primarily focuses on improving the quality-efficiency trade-off of neural information retrieval (IR) systems. Improved quality of IR systems has the potential benefit of improving user experience by improving the quality of IR queries. While search products themselves may have some negative societal impacts, it is unlikely that our work will have any direct path to negative applications or an affect on these impacts. As for limitations, a primary limitation \crisp\ is that it fixes the maximum number of clusters $k$ to be used during training in advance. This prevents the model from learning the optimal number of clusters to use for a given query or document, which is a benefit of unpruned multi-vector models. We leave the exploration of methods which adapt the number of clusters, as well as the exploration of alternative clustering mechanisms, to future work.

\newpage

\bibliography{neurips_2025, jayaram}

\begin{thebibliography}{10}

\bibitem{clavie2024reducingfootprintmultivectorretrieval}
B.~Clavié, A.~Chaffin, and G.~Adams.
\newblock Reducing the footprint of multi-vector retrieval with minimal performance impact via token pooling, 2024.

\bibitem{dhulipala2024muveramultivectorretrievalfixed}
L.~Dhulipala, M.~Hadian, R.~Jayaram, J.~Lee, and V.~Mirrokni.
\newblock Muvera: Multi-vector retrieval via fixed dimensional encodings, 2024.

\bibitem{engels2024dessert}
J.~Engels, B.~Coleman, V.~Lakshman, and A.~Shrivastava.
\newblock Dessert: An efficient algorithm for vector set search with vector set queries.
\newblock {\em Advances in Neural Information Processing Systems}, 36, 2024.

\bibitem{formal2021white}
T.~Formal, B.~Piwowarski, and S.~Clinchant.
\newblock A white box analysis of colbert.
\newblock In {\em Advances in Information Retrieval: 43rd European Conference on IR Research, ECIR 2021, Virtual Event, March 28--April 1, 2021, Proceedings, Part II 43}, pages 257--263. Springer, 2021.

\bibitem{formal2022match}
T.~Formal, B.~Piwowarski, and S.~Clinchant.
\newblock Match your words! a study of lexical matching in neural information retrieval.
\newblock In {\em European Conference on Information Retrieval}, pages 120--127. Springer, 2022.

\bibitem{gao2021coil}
L.~Gao, Z.~Dai, and J.~Callan.
\newblock Coil: Revisit exact lexical match in information retrieval with contextualized inverted list.
\newblock {\em arXiv preprint arXiv:2104.07186}, 2021.

\bibitem{hofstatter2022introducing}
S.~Hofst{\"a}tter, O.~Khattab, S.~Althammer, M.~Sertkan, and A.~Hanbury.
\newblock Introducing neural bag of whole-words with colberter: Contextualized late interactions using enhanced reduction.
\newblock In {\em Proceedings of the 31st ACM International Conference on Information \& Knowledge Management}, pages 737--747, 2022.

\bibitem{khattab2020colbert}
O.~Khattab and M.~Zaharia.
\newblock Colbert: Efficient and effective passage search via contextualized late interaction over bert.
\newblock In {\em Proceedings of the 43rd International ACM SIGIR conference on research and development in Information Retrieval}, pages 39--48, 2020.

\bibitem{lee2024rethinking}
J.~Lee, Z.~Dai, S.~M.~K. Duddu, T.~Lei, I.~Naim, M.-W. Chang, and V.~Zhao.
\newblock Rethinking the role of token retrieval in multi-vector retrieval.
\newblock {\em Advances in Neural Information Processing Systems}, 36, 2024.

\bibitem{li2025making}
C.~Li, M.~Qin, S.~Xiao, J.~Chen, K.~Luo, D.~Lian, Y.~Shao, and Z.~Liu.
\newblock Making text embedders few-shot learners.
\newblock In {\em The Thirteenth International Conference on Learning Representations}, 2025.

\bibitem{lin2024fine}
W.~Lin, J.~Chen, J.~Mei, A.~Coca, and B.~Byrne.
\newblock Fine-grained late-interaction multi-modal retrieval for retrieval augmented visual question answering.
\newblock {\em Advances in Neural Information Processing Systems}, 36, 2024.

\bibitem{liu2024analysis}
Q.~Liu, G.~Guo, J.~Mao, Z.~Dou, J.-R. Wen, H.~Jiang, X.~Zhang, and Z.~Cao.
\newblock An analysis on matching mechanisms and token pruning for late-interaction models.
\newblock {\em ACM Transactions on Information Systems}, 42(5):1--28, 2024.

\bibitem{lupart2023ms}
S.~Lupart, T.~Formal, and S.~Clinchant.
\newblock Ms-shift: An analysis of ms marco distribution shifts on neural retrieval.
\newblock In {\em European Conference on Information Retrieval}, pages 636--652. Springer, 2023.

\bibitem{macavaney2025efficientconstantspacemultivectorretrieval}
S.~MacAvaney, A.~Mallia, and N.~Tonellotto.
\newblock Efficient constant-space multi-vector retrieval, 2025.

\bibitem{macavaney2024reproducibility}
S.~MacAvaney and N.~Tonellotto.
\newblock A reproducibility study of plaid.
\newblock {\em arXiv preprint arXiv:2404.14989}, 2024.

\bibitem{muennighoff2022mteb}
N.~Muennighoff, N.~Tazi, L.~Magne, and N.~Reimers.
\newblock Mteb: Massive text embedding benchmark.
\newblock {\em arXiv preprint arXiv:2210.07316}, 2022.

\bibitem{nguyen2016ms}
T.~Nguyen, M.~Rosenberg, X.~Song, J.~Gao, S.~Tiwary, R.~Majumder, and L.~Deng.
\newblock Ms marco: A human-generated machine reading comprehension dataset.
\newblock 2016.

\bibitem{Ni2022LargeDE}
J.~Ni, C.~Qu, J.~Lu, Z.~Dai, G.~H. {\'{A}}brego, J.~Ma, V.~Y. Zhao, Y.~Luan, K.~B. Hall, M.~Chang, and Y.~Yang.
\newblock Large dual encoders are generalizable retrievers.
\newblock In Y.~Goldberg, Z.~Kozareva, and Y.~Zhang, editors, {\em Proceedings of the 2022 Conference on Empirical Methods in Natural Language Processing, {EMNLP} 2022, Abu Dhabi, United Arab Emirates, December 7-11, 2022}, pages 9844--9855. Association for Computational Linguistics, 2022.

\bibitem{qian2023multivector}
Y.~Qian, J.~Lee, S.~M.~K. Duddu, Z.~Dai, S.~Brahma, I.~Naim, T.~Lei, and V.~Y. Zhao.
\newblock Multi-vector retrieval as sparse alignment, 2022.

\bibitem{raffel2020t5}
C.~Raffel, N.~Shazeer, A.~Roberts, K.~Lee, S.~Narang, M.~Matena, Y.~Zhou, W.~Li, and P.~J. Liu.
\newblock Exploring the limits of transfer learning with a unified text-to-text transformer.
\newblock {\em J. Mach. Learn. Res.}, 21(1), Jan. 2020.

\bibitem{santhanam2022plaid}
K.~Santhanam, O.~Khattab, C.~Potts, and M.~Zaharia.
\newblock Plaid: an efficient engine for late interaction retrieval.
\newblock In {\em Proceedings of the 31st ACM International Conference on Information \& Knowledge Management}, pages 1747--1756, 2022.

\bibitem{santhanam-etal-2022-colbertv2}
K.~Santhanam, O.~Khattab, J.~Saad-Falcon, C.~Potts, and M.~Zaharia.
\newblock {C}ol{BERT}v2: Effective and efficient retrieval via lightweight late interaction.
\newblock In {\em Proceedings of the 2022 Conference of the North American Chapter of the Association for Computational Linguistics: Human Language Technologies}, pages 3715--3734, Seattle, United States, July 2022. Association for Computational Linguistics.

\bibitem{gemmateam2024gemmaopenmodelsbased}
G.~Team, T.~Mesnard, C.~Hardin, R.~Dadashi, S.~Bhupatiraju, S.~Pathak, L.~Sifre, M.~Rivière, M.~S. Kale, J.~Love, P.~Tafti, L.~Hussenot, P.~G. Sessa, A.~Chowdhery, A.~Roberts, A.~Barua, A.~Botev, A.~Castro-Ros, A.~Slone, A.~Héliou, A.~Tacchetti, A.~Bulanova, A.~Paterson, B.~Tsai, B.~Shahriari, C.~L. Lan, C.~A. Choquette-Choo, C.~Crepy, D.~Cer, D.~Ippolito, D.~Reid, E.~Buchatskaya, E.~Ni, E.~Noland, G.~Yan, G.~Tucker, G.-C. Muraru, G.~Rozhdestvenskiy, H.~Michalewski, I.~Tenney, I.~Grishchenko, J.~Austin, J.~Keeling, J.~Labanowski, J.-B. Lespiau, J.~Stanway, J.~Brennan, J.~Chen, J.~Ferret, J.~Chiu, J.~Mao-Jones, K.~Lee, K.~Yu, K.~Millican, L.~L. Sjoesund, L.~Lee, L.~Dixon, M.~Reid, M.~Mikuła, M.~Wirth, M.~Sharman, N.~Chinaev, N.~Thain, O.~Bachem, O.~Chang, O.~Wahltinez, P.~Bailey, P.~Michel, P.~Yotov, R.~Chaabouni, R.~Comanescu, R.~Jana, R.~Anil, R.~McIlroy, R.~Liu, R.~Mullins, S.~L. Smith, S.~Borgeaud, S.~Girgin, S.~Douglas, S.~Pandya, S.~Shakeri, S.~De, T.~Klimenko, T.~Hennigan, V.~Feinberg, W.~Stokowiec,
  Y.~hui Chen, Z.~Ahmed, Z.~Gong, T.~Warkentin, L.~Peran, M.~Giang, C.~Farabet, O.~Vinyals, J.~Dean, K.~Kavukcuoglu, D.~Hassabis, Z.~Ghahramani, D.~Eck, J.~Barral, F.~Pereira, E.~Collins, A.~Joulin, N.~Fiedel, E.~Senter, A.~Andreev, and K.~Kenealy.
\newblock Gemma: Open models based on gemini research and technology, 2024.

\bibitem{thakur2021beir}
N.~Thakur, N.~Reimers, A.~R{\"u}ckl{\'e}, A.~Srivastava, and I.~Gurevych.
\newblock {BEIR}: A heterogeneous benchmark for zero-shot evaluation of information retrieval models.
\newblock In {\em Thirty-fifth Conference on Neural Information Processing Systems Datasets and Benchmarks Track (Round 2)}, 2021.

\bibitem{wachsmuth2018retrieval}
H.~Wachsmuth, S.~Syed, and B.~Stein.
\newblock Retrieval of the best counterargument without prior topic knowledge.
\newblock In {\em Proceedings of the 56th Annual Meeting of the Association for Computational Linguistics (Volume 1: Long Papers)}, pages 241--251, 2018.

\bibitem{wang2021pseudo}
X.~Wang, C.~Macdonald, N.~Tonellotto, and I.~Ounis.
\newblock Pseudo-relevance feedback for multiple representation dense retrieval.
\newblock In {\em Proceedings of the 2021 ACM SIGIR International Conference on Theory of Information Retrieval}, pages 297--306, 2021.

\bibitem{wang2023reproducibility}
X.~Wang, C.~Macdonald, N.~Tonellotto, and I.~Ounis.
\newblock Reproducibility, replicability, and insights into dense multi-representation retrieval models: from colbert to col.
\newblock In {\em Proceedings of the 46th International ACM SIGIR Conference on Research and Development in Information Retrieval}, pages 2552--2561, 2023.

\bibitem{weller2023nevir}
O.~Weller, D.~Lawrie, and B.~Van~Durme.
\newblock Nevir: Negation in neural information retrieval.
\newblock {\em arXiv preprint arXiv:2305.07614}, 2023.

\bibitem{yao2021filip}
L.~Yao, R.~Huang, L.~Hou, G.~Lu, M.~Niu, H.~Xu, X.~Liang, Z.~Li, X.~Jiang, and C.~Xu.
\newblock Filip: Fine-grained interactive language-image pre-training.
\newblock {\em arXiv preprint arXiv:2111.07783}, 2021.

\bibitem{zhan2022evaluating}
J.~Zhan, X.~Xie, J.~Mao, Y.~Liu, J.~Guo, M.~Zhang, and S.~Ma.
\newblock Evaluating interpolation and extrapolation performance of neural retrieval models.
\newblock In {\em Proceedings of the 31st ACM International Conference on Information \& Knowledge Management}, pages 2486--2496, 2022.

\bibitem{zhang2016neural}
Y.~Zhang, M.~M. Rahman, A.~Braylan, B.~Dang, H.-L. Chang, H.~Kim, Q.~McNamara, A.~Angert, E.~Banner, V.~Khetan, et~al.
\newblock Neural information retrieval: A literature review.
\newblock {\em arXiv preprint arXiv:1611.06792}, 2016.

\end{thebibliography}

\appendix

\newpage

\section{Appendix: Qualitative Clustering Examples}
\label{sec:appendix}
To offer qualitative insights into the behavior of K-means clustering on token embeddings, this appendix presents two examples selected from the ArguAna dataset \cite{wachsmuth2018retrieval}, where the task is to retrieve a counterargument for a given argument query. Each example includes the query text, the corresponding gold counterargument document, with some highlighted tokens that belong to the same cluster. In each case, \crisp\ was able to retrieve the top matching document. The matching highlighted cluster in the document reflects the document cluster that achieved the highest multi-vector similarity (Chamfer) with that particular query cluster. These visual examples highlight a recurring pattern: K-means clustering frequently isolates semantically lighter tokens, such as stopwords, punctuation, and generic terms, into distinct clusters.

% --- Task Description ---
\paragraph{Task Description}
Given a claim, find documents that refute the claim.

% --- Configure tcolorbox styles ---
\tcbset{
    querybox/.style={
        enhanced, % Allows more advanced features
        colback=blue!5!white, % Light blue background
        colframe=blue!60!black, % Darker blue frame
        fonttitle=\bfseries,
        attach boxed title to top left={yshift=-2mm, xshift=3mm},
        boxed title style={colback=blue!60!black, colframe=blue!60!black},
        title=Query Text
    },
    docbox/.style={
        enhanced,
        colback=green!5!white, % Light green background
        colframe=green!60!black, % Darker green frame
        fonttitle=\bfseries,
        attach boxed title to top left={yshift=-2mm, xshift=3mm},
        boxed title style={colback=green!60!black, colframe=green!60!black},
        title=Document Text (Refutation)
    }
}

\subsection{Example 1}

\begin{tcolorbox}[querybox]

\textcolor{white}{\colorbox{black}{Given}} \textcolor{white}{\colorbox{black}{a}} \textcolor{white}{\colorbox{black}{claim}}, \textcolor{white}{\colorbox{black}{find}} \textcolor{black}{\colorbox{white}{documents}} \textcolor{white}{\colorbox{black}{that}} \textcolor{white}{\colorbox{black}{refute}} \textcolor{white}{\colorbox{black}{the}} \textcolor{white}{\colorbox{black}{claim}}. \textcolor{white}{\colorbox{black}{|}} \textcolor{white}{\colorbox{black}{query}}: \textcolor{black}{\colorbox{white}{A}} \textcolor{black}{\colorbox{white}{UN}} \textcolor{black}{\colorbox{white}{standing}} \textcolor{black}{\colorbox{white}{army}} \textcolor{black}{\colorbox{white}{is}} \textcolor{black}{\colorbox{white}{simply}} \textcolor{black}{\colorbox{white}{impossible}} \textcolor{black}{\colorbox{white}{to}} \textcolor{black}{\colorbox{white}{form}}\textcolor{white}{\colorbox{black}{.}}\textcolor{white}{\colorbox{black}{A}} \textcolor{black}{\colorbox{white}{standing}} \textcolor{black}{\colorbox{white}{army}} \textcolor{black}{\colorbox{white}{for}} \textcolor{black}{\colorbox{white}{the}} \textcolor{black}{\colorbox{white}{United}} \textcolor{black}{\colorbox{white}{Nations}} \textcolor{black}{\colorbox{white}{has}} \textcolor{black}{\colorbox{white}{an}} \textcolor{black}{\colorbox{white}{existing}} \textcolor{black}{\colorbox{white}{legal}} \textcolor{black}{\colorbox{white}{framework}}\textcolor{white}{\colorbox{black}{;}} \textcolor{black}{\colorbox{white}{it}} \textcolor{black}{\colorbox{white}{has}} \textcolor{black}{\colorbox{white}{never}} \textcolor{black}{\colorbox{white}{been}} \textcolor{black}{\colorbox{white}{attempted}} \textcolor{black}{\colorbox{white}{in}} \textcolor{black}{\colorbox{white}{practice}} \textcolor{white}{\colorbox{black}{because}} \textcolor{black}{\colorbox{white}{it}} \textcolor{black}{\colorbox{white}{would}} \textcolor{black}{\colorbox{white}{be}} \textcolor{black}{\colorbox{white}{impossible}} \textcolor{black}{\colorbox{white}{to}} \textcolor{black}{\colorbox{white}{create}}. \textcolor{black}{\colorbox{white}{Article}} \textcolor{black}{\colorbox{white}{4}} \textcolor{black}{\colorbox{white}{3}} \textcolor{black}{\colorbox{white}{of}} \textcolor{black}{\colorbox{white}{the}} \textcolor{black}{\colorbox{white}{original}} \textcolor{black}{\colorbox{white}{UN}} \textcolor{black}{\colorbox{white}{Charter}} \textcolor{black}{\colorbox{white}{specifies}} \textcolor{black}{\colorbox{white}{that}} \textcolor{black}{\colorbox{white}{all}} \textcolor{black}{\colorbox{white}{member}} \textcolor{black}{\colorbox{white}{states}} \textcolor{black}{\colorbox{white}{are}} \textcolor{black}{\colorbox{white}{expected}}, \textcolor{black}{\colorbox{white}{upon}} \textcolor{black}{\colorbox{white}{the}} \textcolor{black}{\colorbox{white}{signing}} \textcolor{black}{\colorbox{white}{of}} \textcolor{black}{\colorbox{white}{a}} \textcolor{black}{\colorbox{white}{future}} \textcolor{black}{\colorbox{white}{UN}} \textcolor{black}{\colorbox{white}{agreement}}, \textcolor{black}{\colorbox{white}{to}} \textcolor{black}{\colorbox{white}{provide}} \textcolor{black}{\colorbox{white}{‘forces}}, \textcolor{black}{\colorbox{white}{assistance}} \textcolor{black}{\colorbox{white}{and}} \textcolor{black}{\colorbox{white}{facilities}} \textcolor{black}{\colorbox{white}{’}} \textcolor{black}{\colorbox{white}{for}} \textcolor{black}{\colorbox{white}{the}} \textcolor{black}{\colorbox{white}{maintenance}} \textcolor{black}{\colorbox{white}{of}} \textcolor{black}{\colorbox{white}{international}} \textcolor{black}{\colorbox{white}{peace}} \textcolor{black}{\colorbox{white}{and}} \textcolor{black}{\colorbox{white}{security}}\textcolor{white}{\colorbox{black}{1}}. \textcolor{white}{\colorbox{black}{That}} \textcolor{black}{\colorbox{white}{it}} \textcolor{black}{\colorbox{white}{is}} \textcolor{black}{\colorbox{white}{has}} \textcolor{black}{\colorbox{white}{never}} \textcolor{black}{\colorbox{white}{been}} \textcolor{black}{\colorbox{white}{attempted}} \textcolor{black}{\colorbox{white}{is}} \textcolor{white}{\colorbox{black}{the}} \textcolor{black}{\colorbox{white}{direct}} \textcolor{black}{\colorbox{white}{result}} \textcolor{white}{\colorbox{black}{of}} \textcolor{black}{\colorbox{white}{its}} \textcolor{black}{\colorbox{white}{sheer}} \textcolor{black}{\colorbox{white}{impracticality}}\textcolor{white}{\colorbox{black}{;}} \textcolor{black}{\colorbox{white}{who}} \textcolor{black}{\colorbox{white}{would}} \textcolor{black}{\colorbox{white}{contribute}} \textcolor{black}{\colorbox{white}{the}} \textcolor{black}{\colorbox{white}{troops}}\textcolor{white}{\colorbox{black}{?}} \textcolor{white}{\colorbox{black}{How}} \textcolor{white}{\colorbox{black}{would}} \textcolor{black}{\colorbox{white}{they}} \textcolor{black}{\colorbox{white}{be}} \textcolor{black}{\colorbox{white}{trained}}, \textcolor{black}{\colorbox{white}{and}} \textcolor{black}{\colorbox{white}{ensure}} \textcolor{black}{\colorbox{white}{that}} \textcolor{black}{\colorbox{white}{troops}} \textcolor{black}{\colorbox{white}{trained}} \textcolor{black}{\colorbox{white}{in}} \textcolor{black}{\colorbox{white}{one}} \textcolor{black}{\colorbox{white}{state}} \textcolor{black}{\colorbox{white}{would}} \textcolor{black}{\colorbox{white}{not}} \textcolor{black}{\colorbox{white}{be}} \textcolor{black}{\colorbox{white}{asked}} \textcolor{black}{\colorbox{white}{to}} \textcolor{black}{\colorbox{white}{thereafter}} \textcolor{black}{\colorbox{white}{fire}} \textcolor{black}{\colorbox{white}{on}} \textcolor{black}{\colorbox{white}{their}} \textcolor{black}{\colorbox{white}{own}} \textcolor{black}{\colorbox{white}{colleagues}}\textcolor{white}{\colorbox{black}{?}} \textcolor{white}{\colorbox{black}{Furthermore}}, \textcolor{black}{\colorbox{white}{where}} \textcolor{black}{\colorbox{white}{would}} \textcolor{black}{\colorbox{white}{the}} \textcolor{black}{\colorbox{white}{U}}\textcolor{white}{\colorbox{black}{.}}N. \textcolor{black}{\colorbox{white}{standing}} \textcolor{black}{\colorbox{white}{army}} \textcolor{black}{\colorbox{white}{be}} \textcolor{black}{\colorbox{white}{located}}, \textcolor{black}{\colorbox{white}{for}} \textcolor{white}{\colorbox{black}{the}} \textcolor{black}{\colorbox{white}{United}} \textcolor{black}{\colorbox{white}{Nations}} \textcolor{black}{\colorbox{white}{has}} \textcolor{black}{\colorbox{white}{no}} \textcolor{black}{\colorbox{white}{land}}\textcolor{black}{\colorbox{white}{,}} \textcolor{black}{\colorbox{white}{and}} \textcolor{white}{\colorbox{black}{the}} \textcolor{black}{\colorbox{white}{United}} \textcolor{black}{\colorbox{white}{States}} \textcolor{black}{\colorbox{white}{would}} \textcolor{black}{\colorbox{white}{not}} \textcolor{black}{\colorbox{white}{take}} \textcolor{black}{\colorbox{white}{kindly}} \textcolor{black}{\colorbox{white}{to}} \textcolor{black}{\colorbox{white}{a}} \textcolor{black}{\colorbox{white}{reprisal}} \textcolor{black}{\colorbox{white}{attack}} \textcolor{black}{\colorbox{white}{on}} \textcolor{black}{\colorbox{white}{the}} \textcolor{black}{\colorbox{white}{UN}} \textcolor{black}{\colorbox{white}{Army}} \textcolor{black}{\colorbox{white}{at}} \textcolor{black}{\colorbox{white}{the}} \textcolor{black}{\colorbox{white}{United}} \textcolor{black}{\colorbox{white}{Nations}} \textcolor{black}{\colorbox{white}{Headquarters}}\textcolor{white}{\colorbox{black}{.}}\textcolor{white}{\colorbox{black}{And}} \textcolor{black}{\colorbox{white}{who}} \textcolor{black}{\colorbox{white}{would}} \textcolor{black}{\colorbox{white}{fund}} \textcolor{black}{\colorbox{white}{this}} \textcolor{black}{\colorbox{white}{army}}\textcolor{white}{\colorbox{black}{?}} \textcolor{white}{\colorbox{black}{The}} \textcolor{black}{\colorbox{white}{United}} \textcolor{black}{\colorbox{white}{States}} \textcolor{black}{\colorbox{white}{hasn}}\textcolor{white}{\colorbox{black}{’}}\textcolor{black}{\colorbox{white}{t}} \textcolor{black}{\colorbox{white}{paid}} \textcolor{black}{\colorbox{white}{its}} \textcolor{black}{\colorbox{white}{bills}} \textcolor{black}{\colorbox{white}{to}} \textcolor{black}{\colorbox{white}{the}} \textcolor{black}{\colorbox{white}{United}} \textcolor{black}{\colorbox{white}{Nations}} \textcolor{black}{\colorbox{white}{in}} \textcolor{black}{\colorbox{white}{years}} \textcolor{black}{\colorbox{white}{due}} \textcolor{black}{\colorbox{white}{to}} \textcolor{black}{\colorbox{white}{their}} \textcolor{black}{\colorbox{white}{opposition}} \textcolor{black}{\colorbox{white}{to}} \textcolor{black}{\colorbox{white}{some}} \textcolor{black}{\colorbox{white}{of}} \textcolor{black}{\colorbox{white}{its}} \textcolor{black}{\colorbox{white}{actions}}\textcolor{white}{\colorbox{black}{/}}\textcolor{white}{\colorbox{black}{What}} \textcolor{black}{\colorbox{white}{is}} \textcolor{black}{\colorbox{white}{there}} \textcolor{black}{\colorbox{white}{in}} \textcolor{black}{\colorbox{white}{place}} \textcolor{black}{\colorbox{white}{to}} \textcolor{black}{\colorbox{white}{prevent}} \textcolor{black}{\colorbox{white}{that}} \textcolor{black}{\colorbox{white}{continuing}}\textcolor{white}{\colorbox{black}{?}}\textcolor{white}{\colorbox{black}{Lastly}}, \textcolor{white}{\colorbox{black}{and}} \textcolor{black}{\colorbox{white}{most}} \textcolor{black}{\colorbox{white}{importantly}}\textcolor{white}{\colorbox{black}{,}} \textcolor{black}{\colorbox{white}{whose}} \textcolor{black}{\colorbox{white}{will}} \textcolor{black}{\colorbox{white}{would}} \textcolor{black}{\colorbox{white}{they}} \textcolor{black}{\colorbox{white}{be}} \textcolor{black}{\colorbox{white}{implementing}}, \textcolor{black}{\colorbox{white}{for}} \textcolor{black}{\colorbox{white}{the}} \textcolor{black}{\colorbox{white}{United}} \textcolor{black}{\colorbox{white}{Nations}} \textcolor{black}{\colorbox{white}{is}} \textcolor{black}{\colorbox{white}{not}} \textcolor{black}{\colorbox{white}{a}} \textcolor{black}{\colorbox{white}{single}} \textcolor{black}{\colorbox{white}{voice}} \textcolor{black}{\colorbox{white}{but}} \textcolor{black}{\colorbox{white}{the}} \textcolor{black}{\colorbox{white}{aggregated}} \textcolor{black}{\colorbox{white}{noise}} \textcolor{black}{\colorbox{white}{of}} \textcolor{black}{\colorbox{white}{its}} \textcolor{black}{\colorbox{white}{member}} \textcolor{black}{\colorbox{white}{states}}\textcolor{white}{\colorbox{black}{?}} \textcolor{white}{\colorbox{black}{The}} \textcolor{black}{\colorbox{white}{Security}} \textcolor{black}{\colorbox{white}{Council}}, \textcolor{black}{\colorbox{white}{which}} \textcolor{black}{\colorbox{white}{currently}} \textcolor{black}{\colorbox{white}{dictates}} \textcolor{black}{\colorbox{white}{the}} \textcolor{black}{\colorbox{white}{form}} \textcolor{black}{\colorbox{white}{that}} \textcolor{black}{\colorbox{white}{U}}\textcolor{white}{\colorbox{black}{.}}N. \textcolor{black}{\colorbox{white}{peacekeeping}} \textcolor{black}{\colorbox{white}{operations}} \textcolor{black}{\colorbox{white}{take}}\textcolor{black}{\colorbox{white}{,}} \textcolor{black}{\colorbox{white}{are}} \textcolor{black}{\colorbox{white}{not}} \textcolor{black}{\colorbox{white}{a}} \textcolor{black}{\colorbox{white}{group}} \textcolor{black}{\colorbox{white}{to}} \textcolor{black}{\colorbox{white}{whom}} \textcolor{black}{\colorbox{white}{impartiality}} \textcolor{black}{\colorbox{white}{can}} \textcolor{black}{\colorbox{white}{be}} \textcolor{black}{\colorbox{white}{attributed}}\textcolor{white}{\colorbox{black}{.}}\textcolor{white}{\colorbox{black}{A}} \textcolor{black}{\colorbox{white}{U}}\textcolor{black}{\colorbox{white}{.}}\textcolor{black}{\colorbox{white}{N}}\textcolor{black}{\colorbox{white}{.}} \textcolor{black}{\colorbox{white}{standing}} \textcolor{black}{\colorbox{white}{army}} \textcolor{black}{\colorbox{white}{at}} \textcolor{black}{\colorbox{white}{the}} \textcolor{black}{\colorbox{white}{be}} \textcolor{black}{\colorbox{white}{hest}} \textcolor{black}{\colorbox{white}{of}} \textcolor{black}{\colorbox{white}{the}} \textcolor{black}{\colorbox{white}{Security}} \textcolor{black}{\colorbox{white}{Council}} \textcolor{black}{\colorbox{white}{would}} \textcolor{black}{\colorbox{white}{be}} \textcolor{black}{\colorbox{white}{used}} \textcolor{black}{\colorbox{white}{sparingly}} \textcolor{black}{\colorbox{white}{at}} \textcolor{black}{\colorbox{white}{best}} \textcolor{black}{\colorbox{white}{and}} \textcolor{black}{\colorbox{white}{only}} \textcolor{black}{\colorbox{white}{in}} \textcolor{black}{\colorbox{white}{regions}} \textcolor{black}{\colorbox{white}{and}} \textcolor{black}{\colorbox{white}{conflicts}} \textcolor{black}{\colorbox{white}{for}} \textcolor{black}{\colorbox{white}{whom}} \textcolor{black}{\colorbox{white}{all}} \textcolor{black}{\colorbox{white}{the}} \textcolor{black}{\colorbox{white}{P5}} \textcolor{black}{\colorbox{white}{had}} \textcolor{black}{\colorbox{white}{a}} \textcolor{black}{\colorbox{white}{vested}} \textcolor{black}{\colorbox{white}{interest}} \textcolor{black}{\colorbox{white}{in}} \textcolor{black}{\colorbox{white}{the}} \textcolor{black}{\colorbox{white}{maintenance}} \textcolor{black}{\colorbox{white}{of}} \textcolor{black}{\colorbox{white}{peace}}\textcolor{white}{\colorbox{black}{.}}\textcolor{white}{\colorbox{black}{Any}} \textcolor{black}{\colorbox{white}{impartiality}} \textcolor{black}{\colorbox{white}{that}} \textcolor{black}{\colorbox{white}{the}} \textcolor{black}{\colorbox{white}{U}}\textcolor{black}{\colorbox{white}{.}}\textcolor{black}{\colorbox{white}{N}}\textcolor{black}{\colorbox{white}{.}} \textcolor{black}{\colorbox{white}{standing}} \textcolor{black}{\colorbox{white}{army}} \textcolor{black}{\colorbox{white}{had}} \textcolor{black}{\colorbox{white}{in}} \textcolor{black}{\colorbox{white}{theory}} \textcolor{black}{\colorbox{white}{would}} \textcolor{black}{\colorbox{white}{be}} \textcolor{black}{\colorbox{white}{lost}} \textcolor{black}{\colorbox{white}{in}} \textcolor{black}{\colorbox{white}{practice}}\textcolor{white}{\colorbox{black}{.}}\textcolor{white}{\colorbox{black}{1}}. \textcolor{white}{\colorbox{black}{U}}\textcolor{white}{\colorbox{black}{.}}N. \textcolor{black}{\colorbox{white}{Charter}}\textcolor{white}{\colorbox{black}{,}} (\textcolor{white}{\colorbox{black}{1}}945)

\end{tcolorbox}

\begin{tcolorbox}[docbox]

\textcolor{black}{\colorbox{white}{global}} \textcolor{black}{\colorbox{white}{politics}} \textcolor{black}{\colorbox{white}{defence}} \textcolor{black}{\colorbox{white}{war}} \textcolor{black}{\colorbox{white}{peace}} \textcolor{white}{\colorbox{black}{house}} \textcolor{white}{\colorbox{black}{would}} \textcolor{black}{\colorbox{white}{create}} \textcolor{black}{\colorbox{white}{UN}} \textcolor{black}{\colorbox{white}{standing}} \textcolor{black}{\colorbox{white}{army}} \textcolor{white}{\colorbox{black}{A}} \textcolor{white}{\colorbox{black}{U}}\textcolor{white}{\colorbox{black}{.}} \textcolor{black}{\colorbox{white}{N}}\textcolor{white}{\colorbox{black}{.}} \textcolor{black}{\colorbox{white}{standing}} \textcolor{black}{\colorbox{white}{army}} \textcolor{black}{\colorbox{white}{is}} \textcolor{black}{\colorbox{white}{not}} \textcolor{black}{\colorbox{white}{impossible}} \textcolor{black}{\colorbox{white}{to}} \textcolor{black}{\colorbox{white}{form}}\textcolor{white}{\colorbox{black}{.}} \textcolor{black}{\colorbox{white}{The}} \textcolor{black}{\colorbox{white}{United}} \textcolor{black}{\colorbox{white}{Nations}} \textcolor{black}{\colorbox{white}{has}} \textcolor{black}{\colorbox{white}{already}} \textcolor{black}{\colorbox{white}{conclusively}} \textcolor{black}{\colorbox{white}{proved}}, \textcolor{black}{\colorbox{white}{in}} \textcolor{black}{\colorbox{white}{numerous}} \textcolor{black}{\colorbox{white}{peacekeeping}} \textcolor{black}{\colorbox{white}{among}} \textcolor{black}{\colorbox{white}{other}} \textcolor{black}{\colorbox{white}{missions}}, \textcolor{black}{\colorbox{white}{its}} \textcolor{black}{\colorbox{white}{ability}} \textcolor{black}{\colorbox{white}{to}} \textcolor{black}{\colorbox{white}{play}} \textcolor{black}{\colorbox{white}{a}} \textcolor{black}{\colorbox{white}{constructive}}, \textcolor{black}{\colorbox{white}{effective}} \textcolor{black}{\colorbox{white}{military}} \textcolor{black}{\colorbox{white}{role}} \textcolor{black}{\colorbox{white}{in}} \textcolor{black}{\colorbox{white}{interventions}}\textcolor{white}{\colorbox{black}{;}} \textcolor{black}{\colorbox{white}{a}} \textcolor{black}{\colorbox{white}{standing}} \textcolor{black}{\colorbox{white}{army}} \textcolor{black}{\colorbox{white}{would}} \textcolor{black}{\colorbox{white}{merely}} \textcolor{black}{\colorbox{white}{replace}} \textcolor{black}{\colorbox{white}{the}} \textcolor{black}{\colorbox{white}{top}} \textcolor{black}{\colorbox{white}{level}} \textcolor{black}{\colorbox{white}{of}} \textcolor{black}{\colorbox{white}{command}}\textcolor{white}{\colorbox{black}{.}} \textcolor{white}{\colorbox{black}{Instead}} \textcolor{black}{\colorbox{white}{of}} \textcolor{black}{\colorbox{white}{taking}} \textcolor{black}{\colorbox{white}{orders}} \textcolor{black}{\colorbox{white}{from}} \textcolor{black}{\colorbox{white}{the}} \textcolor{black}{\colorbox{white}{top}} \textcolor{black}{\colorbox{white}{brass}} \textcolor{black}{\colorbox{white}{in}} \textcolor{black}{\colorbox{white}{a}} \textcolor{black}{\colorbox{white}{national}} \textcolor{black}{\colorbox{white}{military}}, \textcolor{black}{\colorbox{white}{the}} \textcolor{black}{\colorbox{white}{orders}} \textcolor{black}{\colorbox{white}{would}} \textcolor{black}{\colorbox{white}{come}} \textcolor{black}{\colorbox{white}{from}} \textcolor{black}{\colorbox{white}{United}} \textcolor{black}{\colorbox{white}{Nations}} \textcolor{black}{\colorbox{white}{commanders}}\textcolor{white}{\colorbox{black}{.}} \textcolor{black}{\colorbox{white}{For}} \textcolor{black}{\colorbox{white}{soldiers}} \textcolor{black}{\colorbox{white}{trained}} \textcolor{black}{\colorbox{white}{to}} \textcolor{black}{\colorbox{white}{listen}} \textcolor{black}{\colorbox{white}{and}} \textcolor{black}{\colorbox{white}{respond}} \textcolor{black}{\colorbox{white}{to}} \textcolor{black}{\colorbox{white}{commands}}, \textcolor{black}{\colorbox{white}{this}} \textcolor{black}{\colorbox{white}{would}} \textcolor{black}{\colorbox{white}{constitute}} \textcolor{black}{\colorbox{white}{merely}} \textcolor{black}{\colorbox{white}{a}} \textcolor{black}{\colorbox{white}{subtle}} \textcolor{black}{\colorbox{white}{shift}} \textcolor{black}{\colorbox{white}{that}} \textcolor{black}{\colorbox{white}{would}} \textcolor{black}{\colorbox{white}{not}} \textcolor{black}{\colorbox{white}{alter}} \textcolor{black}{\colorbox{white}{their}} \textcolor{black}{\colorbox{white}{operational}} \textcolor{black}{\colorbox{white}{effectiveness}}\textcolor{white}{\colorbox{black}{.}}\textcolor{white}{\colorbox{black}{Furthermore}}, \textcolor{black}{\colorbox{white}{funding}} \textcolor{black}{\colorbox{white}{would}} \textcolor{black}{\colorbox{white}{be}} \textcolor{black}{\colorbox{white}{provided}} \textcolor{black}{\colorbox{white}{through}} \textcolor{black}{\colorbox{white}{similar}} \textcolor{black}{\colorbox{white}{streams}} \textcolor{black}{\colorbox{white}{to}} \textcolor{black}{\colorbox{white}{how}} \textcolor{black}{\colorbox{white}{peacekeeping}} \textcolor{black}{\colorbox{white}{forces}} \textcolor{black}{\colorbox{white}{are}} \textcolor{black}{\colorbox{white}{funded}} \textcolor{black}{\colorbox{white}{contemporaneously}}\textcolor{white}{\colorbox{black}{,;}} \textcolor{white}{\colorbox{black}{however}}, \textcolor{white}{\colorbox{black}{once}} \textcolor{black}{\colorbox{white}{the}} \textcolor{white}{\colorbox{black}{U}}\textcolor{white}{\colorbox{black}{.}} \textcolor{black}{\colorbox{white}{N}}\textcolor{white}{\colorbox{black}{.}} \textcolor{black}{\colorbox{white}{standing}} \textcolor{black}{\colorbox{white}{army}} \textcolor{black}{\colorbox{white}{has}} \textcolor{black}{\colorbox{white}{proved}} \textcolor{black}{\colorbox{white}{itself}} \textcolor{black}{\colorbox{white}{capable}}, \textcolor{black}{\colorbox{white}{funding}} \textcolor{black}{\colorbox{white}{will}} \textcolor{black}{\colorbox{white}{surely}} \textcolor{black}{\colorbox{white}{come}} \textcolor{black}{\colorbox{white}{from}} \textcolor{black}{\colorbox{white}{those}} \textcolor{black}{\colorbox{white}{states}} \textcolor{black}{\colorbox{white}{who}} \textcolor{white}{\colorbox{black}{recognize}} \textcolor{black}{\colorbox{white}{that}} \textcolor{black}{\colorbox{white}{pooling}} \textcolor{black}{\colorbox{white}{resources}} \textcolor{black}{\colorbox{white}{to}} \textcolor{black}{\colorbox{white}{form}} \textcolor{black}{\colorbox{white}{a}} \textcolor{white}{\colorbox{black}{U}}. \textcolor{black}{\colorbox{white}{N}}\textcolor{black}{\colorbox{white}{.}} \textcolor{black}{\colorbox{white}{army}} \textcolor{black}{\colorbox{white}{is}} \textcolor{black}{\colorbox{white}{more}} \textcolor{black}{\colorbox{white}{prudent}} \textcolor{black}{\colorbox{white}{than}} \textcolor{black}{\colorbox{white}{scratching}} \textcolor{black}{\colorbox{white}{together}} \textcolor{black}{\colorbox{white}{a}} \textcolor{black}{\colorbox{white}{under-resourced}}, \textcolor{black}{\colorbox{white}{native}} \textcolor{black}{\colorbox{white}{army}}\textcolor{white}{\colorbox{black}{.}}

\end{tcolorbox}

\subsection{Example 2}

% Define \textbar if not available (e.g., without amsmath)
\providecommand{\textbar}{|}

\begin{tcolorbox}[querybox]
\textcolor{white}{\colorbox{black}{Given}} \textcolor{white}{\colorbox{black}{a}} \textcolor{white}{\colorbox{black}{claim}} \textcolor{white}{\colorbox{black}{,}} \textcolor{white}{\colorbox{black}{find}} \textcolor{white}{\colorbox{black}{documents}} \textcolor{white}{\colorbox{black}{that}} \textcolor{white}{\colorbox{black}{refute}} \textcolor{white}{\colorbox{black}{the}} \textcolor{white}{\colorbox{black}{claim}} \textcolor{black}{\colorbox{white}{.}} \textcolor{black}{\colorbox{white}{|}} \textcolor{black}{\colorbox{white}{query}} \textcolor{black}{\colorbox{white}{:}} \textcolor{black}{\colorbox{white}{The}} \textcolor{black}{\colorbox{white}{law}} \textcolor{black}{\colorbox{white}{is}} \textcolor{black}{\colorbox{white}{hypocritical}} \textcolor{black}{\colorbox{white}{}} \textcolor{black}{\colorbox{white}{In}} \textcolor{black}{\colorbox{white}{most}} \textcolor{black}{\colorbox{white}{countries}} \textcolor{black}{\colorbox{white}{where}} \textcolor{black}{\colorbox{white}{drugs}} \textcolor{black}{\colorbox{white}{are}} \textcolor{black}{\colorbox{white}{illegal}} \textcolor{black}{\colorbox{white}{,}} \textcolor{black}{\colorbox{white}{tobacco}} \textcolor{black}{\colorbox{white}{and}} \textcolor{black}{\colorbox{white}{alcohol}} \textcolor{black}{\colorbox{white}{,}} \textcolor{black}{\colorbox{white}{which}} \textcolor{black}{\colorbox{white}{arguably}} \textcolor{black}{\colorbox{white}{have}} \textcolor{black}{\colorbox{white}{equally}} \textcolor{black}{\colorbox{white}{devastating}} \textcolor{black}{\colorbox{white}{consequences}} \textcolor{black}{\colorbox{white}{in}} \textcolor{black}{\colorbox{white}{society}} \textcolor{black}{\colorbox{white}{,}} \textcolor{black}{\colorbox{white}{are}} \textcolor{black}{\colorbox{white}{legal}} \textcolor{black}{\colorbox{white}{.}} \textcolor{black}{\colorbox{white}{In}} \textcolor{black}{\colorbox{white}{a}} \textcolor{black}{\colorbox{white}{UK}} \textcolor{black}{\colorbox{white}{study}} \textcolor{black}{\colorbox{white}{,}} \textcolor{black}{\colorbox{white}{alcohol}} \textcolor{black}{\colorbox{white}{was}} \textcolor{black}{\colorbox{white}{shown}} \textcolor{black}{\colorbox{white}{to}} \textcolor{black}{\colorbox{white}{have}} \textcolor{black}{\colorbox{white}{the}} \textcolor{black}{\colorbox{white}{worst}} \textcolor{black}{\colorbox{white}{effects}} \textcolor{black}{\colorbox{white}{of}} \textcolor{black}{\colorbox{white}{any}} \textcolor{black}{\colorbox{white}{drug}} \textcolor{black}{\colorbox{white}{,}} \textcolor{black}{\colorbox{white}{yet}} \textcolor{black}{\colorbox{white}{the}} \textcolor{black}{\colorbox{white}{current}} \textcolor{black}{\colorbox{white}{law}} \textcolor{black}{\colorbox{white}{recognises}} \textcolor{black}{\colorbox{white}{that}} \textcolor{black}{\colorbox{white}{people}} \textcolor{black}{\colorbox{white}{should}} \textcolor{black}{\colorbox{white}{be}} \textcolor{black}{\colorbox{white}{able}} \textcolor{black}{\colorbox{white}{to}} \textcolor{black}{\colorbox{white}{choose}} \textcolor{black}{\colorbox{white}{whether}} \textcolor{black}{\colorbox{white}{they}} \textcolor{black}{\colorbox{white}{drink}} \textcolor{black}{\colorbox{white}{or}} \textcolor{black}{\colorbox{white}{not}} \textcolor{black}{\colorbox{white}{.}} \textcolor{black}{\colorbox{white}{[}} \textcolor{black}{\colorbox{white}{1}} \textcolor{black}{\colorbox{white}{]}} \textcolor{black}{\colorbox{white}{The}} \textcolor{black}{\colorbox{white}{same}} \textcolor{black}{\colorbox{white}{should}} \textcolor{black}{\colorbox{white}{be}} \textcolor{black}{\colorbox{white}{true}} \textcolor{black}{\colorbox{white}{of}} \textcolor{black}{\colorbox{white}{drugs}} \textcolor{black}{\colorbox{white}{.}} \textcolor{black}{\colorbox{white}{}} \textcolor{black}{\colorbox{white}{[}} \textcolor{black}{\colorbox{white}{1}} \textcolor{black}{\colorbox{white}{]}} \textcolor{black}{\colorbox{white}{Professor}} \textcolor{black}{\colorbox{white}{David}} \textcolor{black}{\colorbox{white}{Nutt}} \textcolor{black}{\colorbox{white}{,}} \textcolor{black}{\colorbox{white}{‘}} \textcolor{black}{\colorbox{white}{Drug}} \textcolor{black}{\colorbox{white}{Harms}} \textcolor{black}{\colorbox{white}{in}} \textcolor{black}{\colorbox{white}{the}} \textcolor{black}{\colorbox{white}{UK}} \textcolor{black}{\colorbox{white}{:}} \textcolor{black}{\colorbox{white}{a}} \textcolor{black}{\colorbox{white}{multicriteria}} \textcolor{black}{\colorbox{white}{decision}} \textcolor{black}{\colorbox{white}{analysis}} \textcolor{black}{\colorbox{white}{’,}} \textcolor{black}{\colorbox{white}{The}} \textcolor{black}{\colorbox{white}{Lancet}} \textcolor{black}{\colorbox{white}{,}} \textcolor{black}{\colorbox{white}{Vol}} \textcolor{black}{\colorbox{white}{}} \textcolor{black}{\colorbox{white}{376}} \textcolor{black}{\colorbox{white}{,}} \textcolor{black}{\colorbox{white}{Issue}} \textcolor{black}{\colorbox{white}{}} \textcolor{black}{\colorbox{white}{9752}} \textcolor{black}{\colorbox{white}{,}} \textcolor{black}{\colorbox{white}{pp}} \textcolor{black}{\colorbox{white}{.}} \textcolor{black}{\colorbox{white}{}} \textcolor{black}{\colorbox{white}{1558-1565}} \textcolor{black}{\colorbox{white}{,}} \textcolor{black}{\colorbox{white}{}} \textcolor{black}{\colorbox{white}{6th}} \textcolor{black}{\colorbox{white}{November}} \textcolor{black}{\colorbox{white}{}} \textcolor{black}{\colorbox{white}{2010}} \textcolor{black}{\colorbox{white}{,}}
\end{tcolorbox}

\begin{tcolorbox}[docbox]
\textcolor{white}{\colorbox{black}{th}} \textcolor{white}{\colorbox{black}{addiction}} \textcolor{white}{\colorbox{black}{health}} \textcolor{white}{\colorbox{black}{general}} \textcolor{white}{\colorbox{black}{law}} \textcolor{white}{\colorbox{black}{crime}} \textcolor{white}{\colorbox{black}{policing}} \textcolor{black}{\colorbox{white}{house}} \textcolor{black}{\colorbox{white}{supports}} \textcolor{black}{\colorbox{white}{legal}} \textcolor{black}{\colorbox{white}{isation}} \textcolor{black}{\colorbox{white}{drugs}} \textcolor{black}{\colorbox{white}{Perhaps}} \textcolor{black}{\colorbox{white}{alcohol}} \textcolor{black}{\colorbox{white}{and}} \textcolor{black}{\colorbox{white}{tobacco}} \textcolor{black}{\colorbox{white}{should}} \textcolor{black}{\colorbox{white}{also}} \textcolor{black}{\colorbox{white}{be}} \textcolor{black}{\colorbox{white}{illegal}} \textcolor{black}{\colorbox{white}{.}} \textcolor{black}{\colorbox{white}{However}} \textcolor{black}{\colorbox{white}{,}} \textcolor{black}{\colorbox{white}{one}} \textcolor{black}{\colorbox{white}{of}} \textcolor{black}{\colorbox{white}{the}} \textcolor{black}{\colorbox{white}{reasons}} \textcolor{black}{\colorbox{white}{why}} \textcolor{black}{\colorbox{white}{alcohol}} \textcolor{black}{\colorbox{white}{ranks}} \textcolor{black}{\colorbox{white}{so}} \textcolor{black}{\colorbox{white}{badly}} \textcolor{black}{\colorbox{white}{in}} \textcolor{black}{\colorbox{white}{such}} \textcolor{black}{\colorbox{white}{studies}} \textcolor{black}{\colorbox{white}{is}} \textcolor{black}{\colorbox{white}{because}} \textcolor{black}{\colorbox{white}{of}} \textcolor{black}{\colorbox{white}{its}} \textcolor{black}{\colorbox{white}{legality}} \textcolor{black}{\colorbox{white}{;}} \textcolor{black}{\colorbox{white}{if}} \textcolor{black}{\colorbox{white}{other}} \textcolor{black}{\colorbox{white}{drugs}} \textcolor{black}{\colorbox{white}{were}} \textcolor{black}{\colorbox{white}{legal}} \textcolor{black}{\colorbox{white}{,}} \textcolor{black}{\colorbox{white}{we}} \textcolor{black}{\colorbox{white}{would}} \textcolor{black}{\colorbox{white}{see}} \textcolor{black}{\colorbox{white}{their}} \textcolor{black}{\colorbox{white}{usage}} \textcolor{black}{\colorbox{white}{go}} \textcolor{black}{\colorbox{white}{up}} \textcolor{black}{\colorbox{white}{and}} \textcolor{black}{\colorbox{white}{therefore}} \textcolor{black}{\colorbox{white}{the}} \textcolor{black}{\colorbox{white}{negative}} \textcolor{black}{\colorbox{white}{social}} \textcolor{black}{\colorbox{white}{effects}} \textcolor{black}{\colorbox{white}{they}} \textcolor{black}{\colorbox{white}{produce}} \textcolor{black}{\colorbox{white}{rise}} \textcolor{black}{\colorbox{white}{as}} \textcolor{black}{\colorbox{white}{well}} \textcolor{black}{\colorbox{white}{.}}
\end{tcolorbox}

\subsection{Discussion of Clustering Examples}

The examples presented (Example 1, Example 2) provide a qualitative view into the behavior of K-means clustering on token embeddings. A recurring observation is that certain clusters tend to aggregate tokens with lower semantic weight, such as punctuation, common stopwords (e.g., `a', `the', `of', `and', `that', `would'), or formatting elements (e.g., `|', `,', `.', `;', `?'). For instance, in Example 1, the highlighted cluster contains mostly punctuation, single letters, and generic query/instruction tokens like `Given', `find', `query'. In the corresponding Document Text, the matching cluster also consists mostly of stop words and punctuation.

Note that the prompt prefix used in this particular dataset is prepended to the queries, and the document title, consisting of buzz words provided with the dataset, is prepended to the documents. These instruction words are usually grouped together in a single cluster or together with other semantically light tokens (consider the highlighted cluster in Example 2 and the matching document cluster).

This behavior aligns with the hypothesis behind \crisp: that clustering can serve as a denoising mechanism. During the Chamfer similarity (i.e. MaxSim) calculation inherent in multi-vector retrieval, query clusters dominated by these low-content tokens are less likely to find strong matches within the document's token embeddings, or more precisely these clusters will show no strong preference for one document over another, effectively reducing their contribution to the final similarity score. In the document representation, these clusters often group similarly generic tokens. As Chamfer focuses on the \emph{maximum} similarity for each query token (or centroid), document clusters containing only stopwords are less likely to be the maximal match for query clusters containing substantive terms. Consequently, the influence of these ubiquitous but often irrelevant tokens is mitigated in both the query and document representations, potentially leading to a clearer signal for relevance matching based on more meaningful terms, which helps explain why \crisp\ (via these trained clustering configurations) can sometimes outperform the full multi-vector baseline. If we removed these tokens altogether from the similarity computation we would likely find even a further increase in performance.

\section{Appendix: Task Instruction Prefixes for BEIR Datasets}
\label{sec:appendix_prefixes}

The base model used in our experiments, \texttt{Gemma2B}, is primarily instruction-tuned for conversational tasks. To adapt this model effectively for the diverse retrieval tasks within the BEIR benchmark, we found it necessary to prepend task-specific instruction prefixes to the input queries during both training and inference. Preliminary experiments indicated that omitting these instructions significantly degraded retrieval performance, likely due to the model not being optimally conditioned for the specific retrieval goal of each dataset.

The instruction prefixes in Table \ref{tab:beir_prefixes} were used for each corresponding BEIR dataset.

\begin{table}[htbp]
    \centering
    \caption{Task Instruction Prefixes for BEIR Datasets}
    \label{tab:beir_prefixes}
    \begin{tabularx}{\textwidth}{l X}
        \toprule
        \textbf{Dataset} & \textbf{Instruction Prefix} \\
        \midrule
        \texttt{arguana} & \textit{Given a claim, find documents that refute the claim.} \\
        \texttt{climate-fever} & \textit{Given a claim about climate change, retrieve documents that support or refute the claim.} \\
        \texttt{dbpedia-entity} & \textit{Given a query, retrieve relevant entity descriptions from DBPedia.} \\
        \texttt{fever} & \textit{Given a claim, retrieve documents that support or refute the claim.} \\
        \texttt{fiqa} & \textit{Given a financial question, retrieve user replies that best answer the question.} \\
        \texttt{hotpotqa} & \textit{Given a multi-hop question, retrieve documents that can help answer the question.} \\
        \texttt{msmarco} & \textit{Given a web search query, retrieve relevant passages that answer the query.} \\
        \texttt{nfcorpus} & \textit{Given a question, retrieve relevant documents that best answer the question.} \\
        \texttt{nq} & \textit{Given a question, retrieve Wikipedia passages that answer the question.} \\
        \texttt{quora} & \textit{Given a question, retrieve questions that are semantically equivalent to the given question.} \\
        \texttt{scidocs} & \textit{Given a scientific paper title, retrieve paper abstracts that are cited by the given paper.} \\
        \texttt{scifact} & \textit{Given a scientific claim, retrieve documents that support or refute the claim.} \\
        \texttt{trec-covid} & \textit{Given a query, retrieve documents that answer the query.} \\
        \texttt{webis-touche2020} & \textit{Given a question, retrieve detailed and persuasive arguments that answer the question.} \\
        \texttt{cqadupstack} & \textit{Given a question, retrieve detailed question descriptions from Stackexchange that are duplicates to the given question.} \\
        \bottomrule
    \end{tabularx}
\end{table}

Applying these specific instructions helps align the conversational base model with the target retrieval task for each dataset.

\section{Appendix: Dataset Licenses}
\label{sec:appendix_licenses}
In the main body of the paper we cite references for both our training \cite{li2025making} and evaluation \cite{thakur2021beir} datasets. We include the URL where the training datasets are available. Regarding the evaluation datasets, we refer to the licensing information  disclosed in  page 20 of \cite{thakur2021beir}. Regarding the training dataset, our investigation about the licensing of the BGE training data shows the following heterogeneous licensing overview:
    \begin{itemize}
        \item SQuAD: CC BY-SA 4.0.\footnote{https://rajpurkar.github.io/SQuAD-explorer/}
        \item FEVER: Licensed by a combination of Wikipedia Copyright Policy and CC-BY-SA 3.0.\footnote{https://fever.ai/download/fever/license.html}
        \item NQ: Provided under CC BY-SA 3.0 license.\footnote{https://ai.google.com/research/NaturalQuestions/download}
        % \item TREC-NEWS, Robust04, BioASQ: Data collection archives are under Copyright.
        \item NLI - SNLI under CC BY-SA 4.0,\footnote{https://nlp.stanford.edu/projects/snli/} MNLI under a combination of the OANC’s license, CC BY-SA 3.0, CC BY 3.0\footnote{https://huggingface.co/datasets/nyu-mll/multi\_nli} with preparation code licensed under MIT license\footnote{https://github.com/princeton-nlp/SimCSE}
        \item FiQA: Unknown.\footnote{https://sites.google.com/corp/view/fiqa/home}
        \item Emotion-Classification educational and research use only.\footnote{https://github.com/dair-ai/emotion\_dataset}
        \item MTOPIntent-Classification CC BY-SA 4.0.\footnote{https://fb.me/mtop\_dataset link downloads dataset.}
        \item StackOverFlowDupQuestions (LinkSO) Unknown.\footnote{https://sites.google.com/corp/view/linkso}
        \item ArguAna,\footnote{https://zenodo.org/records/3973258} SciDocsRR,\footnote{https://github.com/allenai/scidocs/blob/master/LICENSE} Banking77-Classification\footnote{https://github.com/PolyAI-LDN/task-specific-datasets}: Provided under CC BY 4.0 license.
        \item ArxivClustering annotations licensed under CC0\footnote{https://www.kaggle.com/datasets/Cornell-University/arxiv} with content under one of the following: CC BY-SA 4.0, CC BY-NC-SA 4.0, CC BY-NC-ND 4.0, arXiv.org perpetual, non-exclusive license 1.0, or CC Zero.\footnote{https://info.arxiv.org/help/license/index.html} 
        \item Biorxiv content under one of the following: no reuse/adaptation without permission, CC-BY-NC-ND, CC-BY-ND, CC-BY-NC, CC-BY, or CC0.\footnote{https://www.biorxiv.org/about/FAQ}
        \item Medrxiv content under one of the following: no reuse/adaptation without permission, CC-BY-NC-ND, CC-BY-ND, CC-BY-NC, CC-BY, or CC0.\footnote{https://www.medrxiv.org/about/FAQ}
        \item RedditClustering and StackExchangeClustering Unknown.\footnote{https://github.com/UKPLab/TWEAC-qa-agent-selection} 
        \item TwentyNewsgroupsClustering CC BY 4.0.\footnote{https://archive.ics.uci.edu/dataset/113/twenty+newsgroups}
        \item STS12 Unknown/mixed with some data licensed under Microsoft Research licenses.\footnote{https://web.archive.org/web/20201029123711/https://www.cs.york.ac.uk/semeval-2012/task6/, http://ixa2.si.ehu.eus/stswiki/}
        \item TriviaQA Unknown - passages sourced from web documents and Wikipedia.\footnote{https://nlp.cs.washington.edu/triviaqa/}
        \item AmazonCounterfactualClassification CC BY-SA 4.0.\footnote{https://github.com/amazon-science/amazon-multilingual-counterfactual-dataset}
        \item TweetSentimentExtractionClassification CC BY 4.0.\footnote{https://www.kaggle.com/competitions/tweet-sentiment-extraction/overview}
        \item IMDB-Classification Unknown.\footnote{http://ai.stanford.edu/~amaas/data/sentiment/index.html}
        \item ToxicConversationsClassification cc0-1.0\footnote{https://huggingface.co/datasets/google/jigsaw\_unintended\_bias}
        \item ELI5: Unknown - harvested from Reddit comments.\footnote{https://facebookresearch.github.io/ELI5/}
        % \item SciFact: Provided under the CC BY-NC 2.0 license.
        % \item SCIDOCS: Provided under the GNU General Public License v3.0 license.
        \item HotpotQA: Provided under the CC BY-SA 4.0 license.\footnote{https://hotpotqa.github.io/}
        \item AmazonReviewsClassification: academic research use dataset license.\footnote{https://github.com/awslabs/open-data-docs/tree/main/docs/amazon-reviews-ml}
        \item Quora Duplicate Questions Detection: Unknown.\footnote{https://www.kaggle.com/c/quora-question-pairs}
        \item MSMARCO passage and document ranking is distributed for non-commercial resource purposes by Microsoft, but not extending 
        any license or other intellectual property
        rights.\footnote{https://microsoft.github.io/msmarco/Datasets.html}
        \item STS22: Unknown - harvested news articles.\footnote{https://competitions.codalab.org/competitions/33835}
        \item STSBenchmark: Unknown - aggregation of datasets from SemEval STS shared tasks from 2012-2017.\footnote{http://ixa2.si.ehu.eus/stswiki/} 
    \end{itemize}

\end{document}